\documentclass[preprint,12pt]{elsarticle}
\usepackage{lineno}

\usepackage{amssymb}

\usepackage{color}
\usepackage{xcolor}
\usepackage{url}
\usepackage{booktabs}
\usepackage{rotating}
\usepackage{pifont}
\usepackage{pgfplots}
\pgfplotsset{compat=1.12}
\usepackage{pgf-pie}  
\newcommand*\rot{\rotatebox{90}}
\newcommand*\OK{\ding{51}}

\newlength\MAX  
\newcommand*\Chart[1]{#1~\rlap{\textcolor{black!20}{\rule{\MAX}{2ex}}}\rule{#1\MAX}{2ex}}
\newcommand*\PieChart[1]{\begin{tikzpicture}[xscale=0.05, yscale=0.05] \tikzset{lines/.style={draw=none},}\pie[color={black, black!10}, style={lines}, rotate=90,/tikz/nodes={opacity=0,overlay}]{#1, 100-#1} \end{tikzpicture}}
\journal{Information and Software Technology}

\begin{document}

\begin{frontmatter}


\ead{terence.wong@adelaide.edu.au, markus.wagner@adelaide.edu.au, christoph.treude@unimelb.edu.au}

\title{Self-Adaptive Systems: A Systematic Literature Review Across Categories and Domains}



\author[affiliation1]{Terence Wong}

\author[affiliation1]{Markus Wagner}

\author[affiliation2]{Christoph Treude}

\address[affiliation1]
{%
School of Computer Science, The University of Adelaide, Australia
}
\address[affiliation2]
{%
School of Computing and Information Systems, The University of Melbourne, Australia
}


\begin{abstract}

Context: Championed by IBM’s vision of autonomic computing paper in 2003, the autonomic computing research field has seen increased research activity over the last 20 years. Several conferences (SEAMS, SASO, ICAC) and workshops (SISSY) have been established and have contributed to the autonomic computing knowledge base in search of a new kind of system -- a self-adaptive system (SAS). These systems are characterized by being context-aware and can act on that awareness. The actions carried out could be on the system or on the context (or environment). The underlying goal of a SAS is the sustained achievement of its goals despite changes in its environment.

Objective: Despite a number of literature reviews on specific aspects of SASs ranging from their requirements to quality attributes, we lack a systematic understanding of the current state of the art. 

Method: This paper contributes a systematic literature review into self-adaptive systems using the dblp computer science bibliography as a database. We filtered the records systematically in successive steps to arrive at 293 relevant papers. Each paper was critically analyzed and categorized into an attribute matrix. This matrix consisted of five categories, with each category having multiple attributes. The attributes of each paper, along with the summary of its contents formed the basis of the literature review that spanned 30 years (1990-2020). 

Results: We characterize the maturation process of the research area from theoretical papers over practical implementations to more holistic and generic approaches, frameworks, and exemplars, applied to areas such as networking, web services, and robotics, with much of the recent work focusing on IoT and IaaS.

Conclusion: While there is an ebb and flow of application domains, domains like bio-inspired approaches, security, and cyber physical systems showed promise to grow heading into the 2020s.

\end{abstract}




\end{frontmatter}

\sloppy

\section{Introduction}

Information systems and the devices that contain them are growing more complex and more pervasive in society. This is due to the increased quantity of and increased demands on these systems. In the past, these systems operated standalone in isolation from other devices with a narrowly prescribed function (e.g., early mobile phones or PCs).  In the current age, these systems have become more distributed (e.g., cloud servers, sensor networks) and more complex (e.g., current mobile phones or PCs). Users expect these systems to be always connected and highly integrated with minimal downtime.  

As the nature of information systems has rapidly evolved, the original in-isolation requirements of these systems are outdated. It is no longer beneficial to specify a system's behavior at design time because there are overwhelmingly more potential system states than can be designed for. This is due to the increased functionality and connectedness of modern information systems. Hence, it is desirable for a system to be able to adapt its behavior at run time to changes in its context (or environment) to ensure the continual achievement of its goals. Such a system is called a self-adaptive system (SAS)~\cite{Oreizy1998,Kephart2003}. 

In the grand challenge presented by IBM~\cite{Kephart2003}, SASs were to provide self-management properties such as self-configuration, self-optimization, self-healing, and self-protection. This challenge led to the establishment of the International Conference on Autonomic Computing (ICAC) as well as establishing foundational theory on SASs~\cite{1301336}.
Self-adaptive approaches range from static, reactive, parametric solutions to dynamic, proactive, structural solutions. The former approaches are based in predetermined plans and configurations while the latter approaches commonly leverage the power of AI/ML~\cite{Weyns2018}.

While systematic reviews have been conducted to characterize the state of the art of specific aspects of self-adaptive systems such as requirements~\cite{yang2014systematic, sucipto2015systematic}, claims and evidence~\cite{review1102}, quality attributes~\cite{mahdavi2017systematic}, and machine learning in the context of SAS~\cite{saputri2020application}, there is a lack of systematic work spanning and providing an overview of all aspects of self-adaptive systems. In this work, we contribute a systematic literature review into self-adaptive systems which categorizes papers into five categories (Technological, Methodology, Perspective, Analytical, Empirical) and summarizes trends and developments across time, categories, and related attributes.

Our review characterizes the development of a research area over 30 years, from theoretical and model based papers in the 1990s and practical implementations and frameworks in the early 2000s to the ramping up period featuring more holistic and generic approaches, which were forthwith extended to frameworks and exemplars. While web services were prevalent in the field as application domain for much of the foundational years, Internet of Things (IoT) and Infrastructure as a Service (IaaS) have dominated self-adaptive systems in the last five years. 

To the best of our knowledge, this is the first systematic literature review on self-adaptive systems produced after the ramping up of the research field that is not limited to a specific aspect of the field, such as the aforementioned reviews. Our main contribution is the cataloging of research on self-adaptive systems and the organization of this catalog according to paper categories, application domains, and additional attributes specific to each paper category. This catalog serves multiple purposes: (i) summarizing the state of the art for practitioners by providing a multi-faceted overview of previous and current work, (ii) identifying current trends and gaps for researchers by organizing content over time and exposing common and uncommon attribute combinations, and (iii) highlighting areas with high potential impact to guide educators in assessing which skills will be particularly relevant for future decision makers.

We find that the current state of the art in self-adaptive systems is focused on developing methodologies and technology in the area of cloud-based services, such as IoT and IaaS. Even though research on self-adaptive systems tends to be diverse, empirical and analytical research is currently playing a smaller role, as are other application domains. Perhaps encouragingly, many approaches are evaluated using real-world case studies, with less reliance on simulations. The importance of self-adaptive systems is rapidly growing in areas such as bio-inspired approaches, security, and cyber physical systems. Going forward, we expect to see a shift towards empirical studies as the research field continues to mature, with industrial case studies in a diverse range of application domains.

The remainder of this paper is structured as follows: Section~2 presents background and highlights findings from related reviews. Section~3 presents our methodology before we provide a chronological overview of research on self-adaptive systems in Section~4. Section~5 discusses the limitations and threats to validity of this work, before Section~6 concludes this paper.


%
%
%
%


\section{Background and Related Reviews}

In this section, we define key terminology in the area of self-adaptive systems in the context of existing literature reviews which often focus on specific aspects of self-adaptive systems.

\textbf{Autonomic computing} is a self-managing computing model named after, and patterned on, the human body's autonomic nervous system. It deals with the design and the construction of computing systems that possess inherent self-managing capabilities~\cite{chainbi2005applying}. The term gained popularity in the early 2000s as a result of IBM’s autonomic computing initiative. Seminal articles by Ganek, Kephart and others~\cite{Kephart2003, Ganek2003, Ganek2004} describe the fundamental characteristics of autonomic systems, a framework for how systems will evolve to become more self-managing, and the key role for open industry standards needed to support autonomic behavior in heterogeneous system environments. A first consideration of research challenges in the field of autonomic computing was published by Kephart in 2005~\cite{Kephart2005}, with a focus on autonomic element challenges, autonomic system challenges, and human-computer challenges. A survey published in 2008 found autonomicity to be not well defined, leading to different systems adhering to different degrees of autonomicity~\cite{review0605}.

A \textbf{self-adaptive system} is a closed-loop system with a feedback loop aiming to adjust itself to changes during its operation~\cite{review0607}. In one of the few review articles that span the entire field of self-adaptive software, Salehie and Tahvildari present a landscape of research in self-adaptive software by highlighting relevant disciplines and prominent research projects~\cite{review0607}. Since the publication of their review in 2009, other review articles have focused on specific aspects of self-adaptive systems, ranging from engineering approaches for self-adaptive systems~\cite{review1105} and requirements modeling and analysis for self-adaptive systems~\cite{yang2014systematic, sucipto2015systematic} to machine learning in the context of self-adaptive systems~\cite{saputri2020application} and quality attributes that are frequently addressed~\cite{mahdavi2017systematic}.

A \textbf{self-healing system} is a system that is capable of performing a reconfiguration step in order to recover from a permanent fault~\cite{pereira2006performance}. In an early review of work on self-healing systems, Ghosh et al.~\cite{review0603} surveyed research in this field and proposed a strategy of synthesis and classification. This was followed by surveys by Psaier and Dustdar~\cite{review1101} and Schneider et al.~\cite{review1104}. A special case of self-healing systems are \textbf{self-protecting systems}, defined as autonomic systems capable of detecting and mitigating security threats at runtime. Yuan and Malek provided a review of work in this area~\cite{review1103}.

Finally, Weyns et al.~\cite{review1102} provided a review on an aspect that is orthogonal to the types of self-adaptive systems, instead focusing on researchers' claims and supporting evidence in this field. They recommend researchers to make their assessment methods, tools and data publicly available and to improve discussion of limitations~\cite{review1102}.

Contrary to these reviews and more in line with the 2009 review by Salehie and Tahvildari~\cite{review0607}, we do not limit ourselves to a particular aspect of self-adaptive systems or a particular aspect of published research in the area, instead aiming to provide a high-level overview across categories and application domains.

\section{Methodology}\label{sec:method}

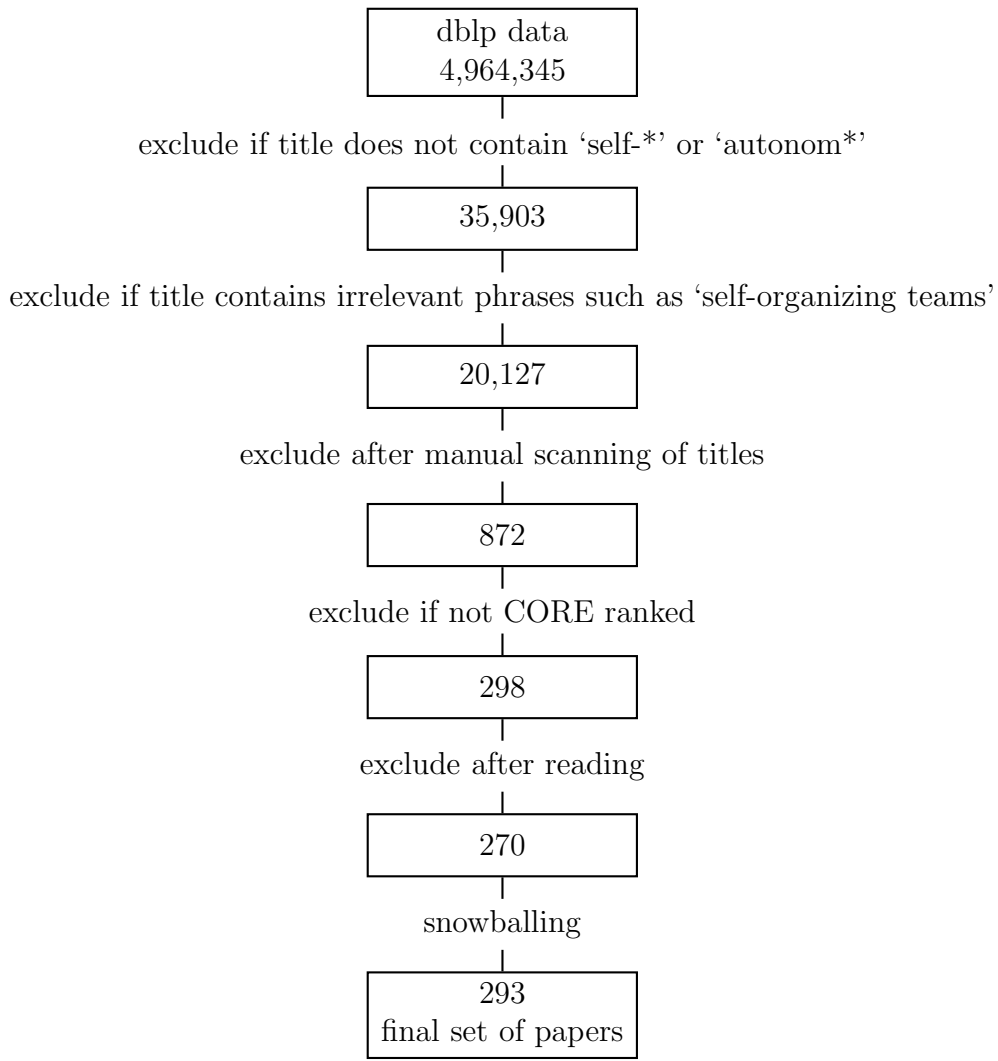
\begin{figure}
\centering
  \begin{tikzpicture}[auto,
    block_center/.style ={rectangle, draw=black, thick, fill=white,
      text width=8em, text centered,
      minimum height=2em},
    block_noborder/.style ={rectangle, draw=none, thick, fill=none,
      text width=\linewidth, text centered, minimum height=1em},
      line/.style ={draw, thick, shorten >=0pt} ]
    \matrix [column sep=5mm,row sep=3mm] {
      \node [block_center] (1) {dblp data\\4,964,345}; \\
      \node [block_noborder] (1_2) {exclude if title does not contain `self-*' or `autonom*'}; \\
      \node [block_center] (2) {35,903}; \\
      \node [block_noborder] (2_3) {exclude if title contains irrelevant phrases such as `self-organizing teams'}; \\
      \node [block_center] (3) {20,127}; \\
      \node [block_noborder] (3_4) {exclude after manual scanning of titles}; \\
      \node [block_center] (4) {872}; \\
      \node [block_noborder] (4_5) {exclude if not CORE ranked}; \\
      \node [block_center] (5) {298}; \\
      \node [block_noborder] (5_6) {exclude after reading}; \\
      \node [block_center] (6) {270}; \\
      \node [block_noborder] (6_7) {snowballing}; \\
      \node [block_center] (7) {293\\final set of papers}; \\
    };
    \begin{scope}[every path/.style=line]
      \path (1) -- (1_2);
      \path (1_2) -- (2);
      \path (2) -- (2_3);
      \path (2_3) -- (3);
      \path (3) -- (3_4);
      \path (3_4) -- (4);
      \path (4) -- (4_5);
      \path (4_5) -- (5);
      \path (5) -- (5_6);
      \path (5_6) -- (6);
      \path (6) -- (6_7);
      \path (6_7) -- (7);
    \end{scope}
  \end{tikzpicture}
\caption{Methodology Overview}
\label{fig:methodology}
\end{figure}

This section outlines the methodology followed in our systematic review, detailing the steps recommended by Kitchenham et al.~\cite{kitchenham2009systematic}.

\subsection{Research Questions}

We used the following research questions to drive our data collection and analysis:

\begin{description}
\item[RQ1] What is the current state of the art in self-adaptive systems?
\item[RQ2] How has the state of the art evolved over time?
\item[RQ3] Which are the application domains of self-adaptive systems over time?
\end{description}

\subsection{Study Protocol}\label{sec:protocol}

To select literature to include in our systematic review on self-adaptive systems, we used the dblp computer science bibliography (dblp) database as a starting point, effectively eliminating non-computer science literature from the process. Dblp contains historical snapshots of the database enabling reproducibility of results. All results in this review were produced using the snapshot file dblp-2020-03-02.xml.\footnote{\url{https://dblp.org/xml/release/dblp-2020-03-02.xml.gz}} 

We chose dblp as the starting point of our analysis since it provides a consistent format for all articles indexed in the database which allowed us to employ consistent search criteria (such as lower-case vs.~upper-case) across content published by different publishers. Dblp's focus on ``major computer science publications''\footnote{\url{https://dblp.org/faq/What+is+dblp.html}} gives us a narrower focus than a more general search engine and a finite number of search results that does not vary on a daily basis, as it would for example on Google Scholar. We acknowledge that using a different database as a starting point would have resulted in a different set of papers, see Section~\ref{sec:threats} for a discussion of the corresponding trade-offs.

The methodology is broken down into five main stages which combine automated and manual processes to keep the required amount of manual work (i.e., reading abstracts and papers) manageable while ensuring the quality of the selection process: Pre-filtering, word frequency filtering, venue selection, abstract reading, and snowballing to reduce a first subset of 35,903 publications to 293 publications. Figure~\ref{fig:methodology} shows a high-level overview of the process.


\subsubsection{Pre-Filtering}

The dblp data containing 4,964,345 papers was filtered to include only journal articles and papers published in conference proceedings, with a page count of over five to focus our analysis on substantial and fully evaluated research contributions which appeared in peer-reviewed venues. Capitalization was removed to assist with the filtering process. This established the base dataset. The base dataset was searched with grep for two terms: self-* and autonom* to produce two separate datasets. These two keywords were deemed as broadly relevant to the study area. The total number of publications matching these keywords was 35,903. 

\subsubsection{Word Frequency Filtering}\label{sec:wordanalysis}

The initial search for publication titles including `self-*' or `autonom*' produced too many papers to manually analyze. However, we noticed that many papers which matched the search could be easily excluded since their topics were clearly out of scope, e.g., self-driving cars. To formalize this process, we further filtered papers based on additional words in the paper titles. We first determined the distribution of words after the keywords self-* and autonom*. Tables~\ref{tab:afterSelf} and~\ref{tab:afterAutonom} show the distribution of the five most frequent words for self-* and for autonom*. 

\begin{table}[t]
\centering
\caption{Words after self-}
\begin{tabular}{lr}
\toprule
Word & Frequency \\
\midrule
organizing & 2,981 \\ 
adaptive & 1,682 \\ 
stabilizing & 632 \\ 
organization & 564 \\ 
organized & 515 \\ 
\bottomrule
\end{tabular}
\label{tab:afterSelf}
\end{table}

\begin{table}[t]
\centering
\caption{Words after autonom*}
\begin{tabular}{lr}
\toprule
Word & Frequency \\
\midrule
vehicles & 576 \\ 
mobile & 554 \\ 
driving & 407 \\ 
systems & 399 \\ 
agents & 399 \\ 
\bottomrule
\end{tabular}
\label{tab:afterAutonom}
\end{table}

Because we observed that self-organizing and self-adaptive frequently co-occurred with terms irrelevant to self-adaptive systems (such as self-organizing teams), we conducted a second word frequency analysis to identify common words appearing after self-organizing and self-adaptive. The top results are shown in Tables~\ref{tab:afterSelfOrganizing} and~\ref{tab:afterSelfAdaptive}.

\begin{table}[t]
\centering
\caption{Words after self-organizing}
\begin{tabular}{lr}
\toprule
Word & Frequency \\
\midrule
maps & 706 \\ 
map & 582 \\ 
neural & 165 \\ 
feature & 109 \\ 
networks & 79 \\ 
\bottomrule
\end{tabular}
\label{tab:afterSelfOrganizing}
\end{table}

\begin{table}[t]
\centering
\caption{Words after self-adaptive}
\begin{tabular}{lr}
\toprule
Word & Frequency \\
\midrule
systems & 159 \\ 
differential & 77 \\ 
software & 76 \\ 
and & 31 \\ 
evolutionary & 27 \\ 
\bottomrule
\end{tabular}
\label{tab:afterSelfAdaptive}
\end{table}


To ensure the validity of deciding which keywords we deemed to be out of scope, two of the authors manually and independently analyzed all words occurring after `self-', `autonom*', `self-organizing', and `self-adaptive' at least 20 times to indicate those that were out of scope (such as self-organizing teams) for exclusion. We calculated inter-rater reliability using Cohen's $\kappa$, see Table~\ref{tab:agreement}. The Cohen’s $\kappa$ value was greater than 0.7 in all cases which was deemed as acceptable. Disagreements were resolved after confirming any ambiguities and biases. This step resulted in a list of words to exclude from the final data set. After omitting papers with the excluded phrases in the title, the number of publications reduced to 20,137.

\begin{table}[t]
\centering
\caption{Inter-rater agreement}
\begin{tabular}{lr}
\toprule
Dataset & Cohen's $\kappa$ \\
\midrule
Autonom* & 0.828 \\ 
Self-* & 0.748 \\ 
Self-organizing * & 1.000 \\ 
Self-adaptive * & 1.000 \\ 
\midrule
50 randomly sampled titles & 0.730 \\ 
\bottomrule
\end{tabular}
\label{tab:agreement}
\end{table}

The titles of the 20,137 publications were manually scanned for relevance, resulting in 872 articles. The purpose of this step was to eliminate papers that matched our keyword filters but were not related to self-adaptive systems. For example, papers on self-driving cars matched our keyword filters but are not related to self-adaptive systems. We again assessed the validity of this step using Cohen's $\kappa$ by having two authors identify relevant papers in a randomly selected subset of 50 papers. The value calculated was 0.73 which was deemed acceptable.  

\subsubsection{Venue Selection}

As a quality gate, we only considered papers published in A*/A conferences or journals, as determined by the CORE ranking.\footnote{The CORE rankings \url{http://portal.core.edu.au/conf-ranks/} and \url{http://portal.core.edu.au/jnl-ranks/} are maintained by the Computing Research and Education Association of Australasia and are used world-wide.} 46\% of all journals and 55\% of all conferences listed by CORE are ranked as A*/A. Papers published in B-ranked venues were included if they came from a journal or conference that was relevant to the study area (e.g., SEAMS\footnote{The CORE rank of SEAMS has changed to A after we conducted this study}, SASO, and TAAS). This step reduced the number of candidate papers to 298. We discuss the trade-offs associated with this step in Section~\ref{sec:threats}.

\subsubsection{Abstract Reading and Attribute Matrix}

The abstracts of the 298 candidate papers were then read 
to categorize each paper into one of five categories based off the paper categorization introduced by the ICSE 2014 conference:\footnote{The categories published in the ICSE 2014 call for papers at \url{https://2014.icse-conferences.org/research} provide one of the most comprehensive categorization schemes for software-related papers.} 

\begin{itemize}
\item Analytical: A paper in which the main contribution relies on new algorithms or mathematical theory.
\item Empirical: A paper in which the main contribution is the empirical study of a software engineering technology or phenomenon.
\item Technological: A paper in which the main contribution is of a technical nature. 
\item Methodological: A paper in which the main contribution is a coherent system of broad principles and practices to interpret or solve a problem. 
\item Perspectives: A paper in which the main contribution is a novel perspective on the field as a whole, or part thereof. 
\end{itemize}

Once each paper was categorized, an attribute matrix for each category was developed. Within each category, the abstract and a pass of the full paper was read to develop an understanding of the common types of attributes for a category. This was done in an iterative approach. 
Once the attribute matrix was finalized, all the papers from each category were categorized using the matrix. This step resulted in 28 papers being discarded as irrelevant. Note that each category has its own attribute matrix since not all attributes apply to all categories. 

We further identified the application domain of each paper, if applicable. If the research has an application focus, the application domain is the business/application sector of the work (e.g. IoT, IaaS, Automotive), otherwise it is the engineering domain (e.g., Web Services, Robotics), or the more general domain (e.g., Bio-inspired, Software Engineering, Security). The application domains are shown in the attribute matrices at the end of the paper and the top 10 most common application domains are summarized in Table~\ref{tab:domains}. Please refer to Tables~\ref{tab:analytical} through~\ref{tab:empirical} for the complete list of papers that we assigned to each application domain.

\subsubsection{Snowballing}

To capture additional papers, snowballing was used on the papers included in the attribute matrix. Where possible, each paper was added to a Scopus list where the references were extracted automatically. This step captured 90\% of the papers in the matrix. Any paper that was referenced more than five times was eligible to be considered. These papers were then filtered through the same criteria as the original papers from the dblp step (i.e., abstracts, CORE rankings). From this step, 23 new papers were added. These papers were categorized into the five categories following the same approach as before. The final counts for each category are shown in Table~\ref{tab:categories} and are considered in the ratios reported in the previous section. Please refer to Tables~\ref{tab:analytical} through~\ref{tab:empirical} for the complete list of papers that we assigned to each category.

\begin{table}[t]
\centering
\caption{Papers per category}
\begin{tabular}{lr}
\toprule
Category & Count \\
\midrule
Technological & 105 \\ 
Methodology & 79 \\ 
Perspective & 51 \\ 
Analytical & 35 \\ 
Empirical & 23 \\ 
\midrule
Total & 293 \\ 
\bottomrule
\end{tabular}
\label{tab:categories}
\end{table}




\begin{table}[t]
\centering
\caption{Papers per application domain}
\begin{tabular}{lr}
\toprule
Application Domain & Count \\
\midrule
Web Services & 48 \\ 
IoT & 36 \\ 
Review & 25 \\ 
Robotics & 23 \\ 
Networking & 19 \\
IaaS & 17 \\ 
Intelligence Surveillance Reconnaissance & 13 \\ 
Software Engineering & 12 \\ 
Automotive & 10 \\ 
Mobile Systems & 10 \\ 
Service-Oriented Systems & 10 \\
\bottomrule
\end{tabular}
\label{tab:domains}
\end{table}


\section{Self-adaptive systems over the years}\label{sec:years}

\begin{figure}
\centering
\begin{tikzpicture}
\begin{axis}[legend style={at={(0,1)},anchor=north west},
symbolic x coords={1990, 1991, 1992, 1993, 1994, 1995, 1996, 1997, 1998, 1999, 2000, 2001, 2002, 2003, 2004, 2005, 2006, 2007, 2008, 2009, 2010, 2011, 2012, 2013, 2014, 2015, 2016, 2017, 2018, 2019, 2020}, xtick=data,
ymin=0,ymax=26, 
x=10,
x tick label style={rotate=25,anchor=north east},
xtick={1990, 1995, ..., 2020},]
\addlegendentry{Publications}
\addplot[smooth,color=black,mark=*,mark options={scale=1.5}] coordinates {(1990,1) (1991,0) (1992,1) (1993,0) (1994,0) (1995,0) (1996,0) (1997,0) (1998,1) (1999,2) (2000,1) (2001,0) (2002,0) (2003,3) (2004,15) (2005,15) (2006,12) (2007,6) (2008,16) (2009,24) (2010,17) (2011,24) (2012,21) (2013,16) (2014,13) (2015,17) (2016,25) (2017,23) (2018,26) (2019,13) (2020,1) };

\end{axis}
\end{tikzpicture}
\caption{Number of publications over time}
\label{fig:papersovertime}
\end{figure}
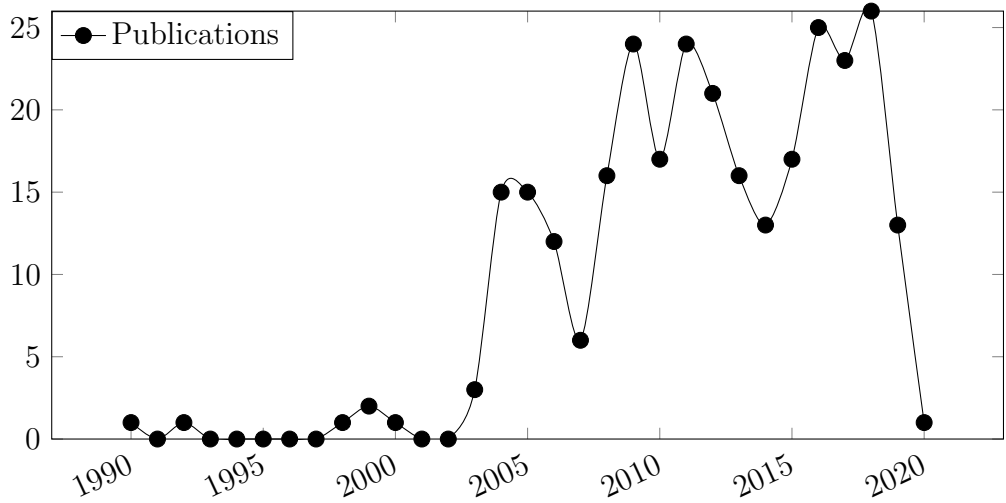

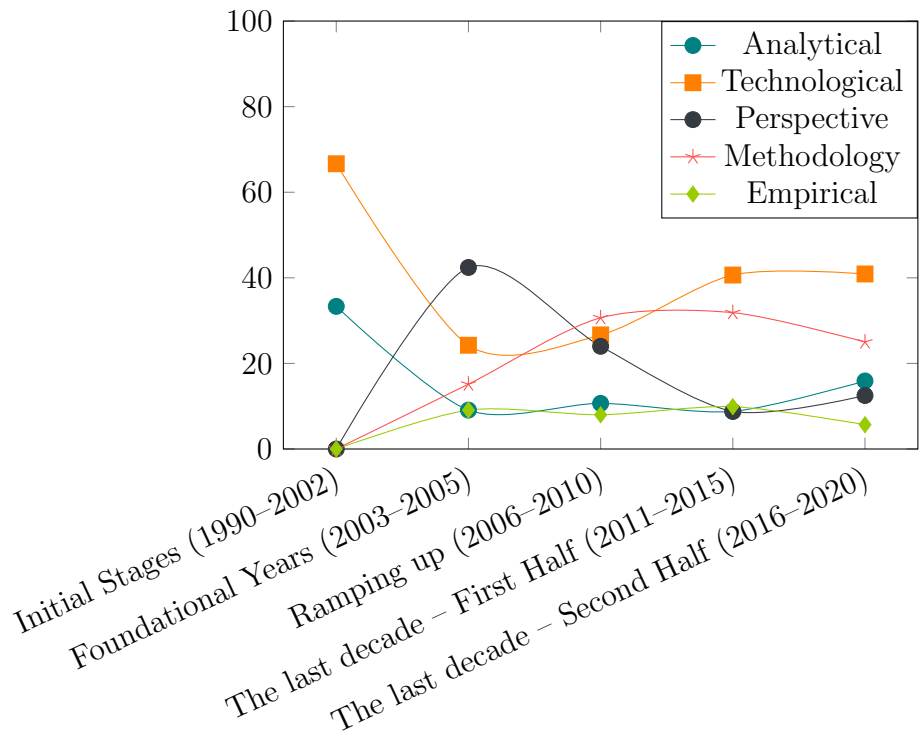
\begin{figure}
\centering
\begin{tikzpicture}
\begin{axis}[legend style={at={(1,1)},anchor=north east},
symbolic x coords={Initial Stages (1990--2002), Foundational Years (2003--2005), Ramping up (2006--2010), The last decade -- First Half (2011--2015), The last decade -- Second Half (2016--2020)}, xtick=data,
ymin=0,ymax=100, 
x=50,
x tick label style={rotate=25,anchor=north east}]
\addlegendentry{Analytical}
\addplot[smooth,color=teal,mark=*,mark options={scale=1.5}] coordinates {(Initial Stages (1990--2002),33.33) (Foundational Years (2003--2005),9.09) (Ramping up (2006--2010),10.66) (The last decade -- First Half (2011--2015),8.79) (The last decade -- Second Half (2016--2020),15.90)
};
\addlegendentry{Technological}
\addplot[smooth,color=orange,mark=square*,mark options={scale=1.5}] coordinates {(Initial Stages (1990--2002),66.66) (Foundational Years (2003--2005),24.24) (Ramping up (2006--2010),26.66) (The last decade -- First Half (2011--2015),40.65) (The last decade -- Second Half (2016--2020),40.90)
};
\addlegendentry{Perspective}
\addplot[smooth,color=cyan!10!black,mark=otimes*,mark options={scale=1.5}] coordinates {(Initial Stages (1990--2002),0.0) (Foundational Years (2003--2005),42.42) (Ramping up (2006--2010),24.00) (The last decade -- First Half (2011--2015),8.79) (The last decade -- Second Half (2016--2020),12.50)
};
\addlegendentry{Methodology}
\addplot[smooth,color=red!70!white,mark=star,mark options={scale=1.5}] coordinates {(Initial Stages (1990--2002),0.0) (Foundational Years (2003--2005),15.15) (Ramping up (2006--2010),30.66) (The last decade -- First Half (2011--2015),31.86) (The last decade -- Second Half (2016--2020),25.0)
};
\addlegendentry{Empirical}
\addplot[smooth,color=lime!80!black,mark=diamond*,mark options={scale=1.5}] coordinates {(Initial Stages (1990--2002),0.0) (Foundational Years (2003--2005),9.09) (Ramping up (2006--2010),8.0) (The last decade -- First Half (2011--2015),9.89) (The last decade -- Second Half (2016--2020),5.68)
};

\end{axis}
\end{tikzpicture}
\caption{Categories over time}
\label{fig:categoriesovertime}
\end{figure}

\begin{figure}
\centering
\begin{tikzpicture}
\begin{axis}[legend style={at={(1,1)},anchor=north east,font=\tiny}, legend columns=2, 
symbolic x coords={Initial Stages (1990--2002), Foundational Years (2003--2005), Ramping up (2006--2010), The last decade -- First Half (2011--2015), The last decade -- Second Half (2016--2020)}, xtick=data,
ymin=0,ymax=100, 
x=50,
x tick label style={rotate=25,anchor=north east}]
\addlegendentry{Other}
\addplot[smooth,color=teal,mark=*,mark options={scale=1.5}] coordinates {(Initial Stages (1990--2002),16.67) (Foundational Years (2003--2005),24.24) (Ramping up (2006--2010),30.67) (The last decade -- First Half (2011--2015),15.38) (The last decade -- Second Half (2016--2020),26.14)};
\addlegendentry{Web Services}
\addplot[smooth,color=orange,mark=square*,mark options={scale=1.5}] coordinates {(Initial Stages (1990--2002),0.00) (Foundational Years (2003--2005),18.18) (Ramping up (2006--2010),18.67) (The last decade -- First Half (2011--2015),20.88) (The last decade -- Second Half (2016--2020),10.23)};
\addlegendentry{IoT}
\addplot[smooth,color=cyan!10!black,mark=otimes*,mark options={scale=1.5}] coordinates {(Initial Stages (1990--2002),16.67) (Foundational Years (2003--2005),9.09) (Ramping up (2006--2010),12.00) (The last decade -- First Half (2011--2015),7.69) (The last decade -- Second Half (2016--2020),18.18)};
\addlegendentry{Review}
\addplot[smooth,color=red!70!white,mark=star,mark options={scale=1.5}] coordinates {(Initial Stages (1990--2002),0.00) (Foundational Years (2003--2005),24.24) (Ramping up (2006--2010),10.67) (The last decade -- First Half (2011--2015),5.49) (The last decade -- Second Half (2016--2020),4.55)};
\addlegendentry{Robotics}
\addplot[smooth,color=lime!80!black,mark=diamond*,mark options={scale=1.5}] coordinates {(Initial Stages (1990--2002),16.67) (Foundational Years (2003--2005),3.03) (Ramping up (2006--2010),9.33) (The last decade -- First Half (2011--2015),12.09) (The last decade -- Second Half (2016--2020),3.41)};
\addlegendentry{Networking}
\addplot[smooth,color=red,mark=*,mark options={scale=1.5}] coordinates {(Initial Stages (1990--2002),33.33) (Foundational Years (2003--2005),6.06) (Ramping up (2006--2010),2.67) (The last decade -- First Half (2011--2015),10.99) (The last decade -- Second Half (2016--2020),3.41)};
\addlegendentry{IaaS}
\addplot[smooth,color=yellow!60!black,mark=square*,mark options={scale=1.5},densely dashed] coordinates {(Initial Stages (1990--2002),0.00) (Foundational Years (2003--2005),3.03) (Ramping up (2006--2010),0.00) (The last decade -- First Half (2011--2015),4.40) (The last decade -- Second Half (2016--2020),13.64)};
\addlegendentry{ISR}
\addplot[smooth,color=black,mark=otimes*,mark options={scale=1.5},densely dashed] coordinates {(Initial Stages (1990--2002),0.00) (Foundational Years (2003--2005),3.03) (Ramping up (2006--2010),2.67) (The last decade -- First Half (2011--2015),6.59) (The last decade -- Second Half (2016--2020),4.55)};
\addlegendentry{Software Engineering}
\addplot[smooth,color=blue,mark=star,mark options={scale=1.5},densely dashed] coordinates {(Initial Stages (1990--2002),16.67) (Foundational Years (2003--2005),6.06) (Ramping up (2006--2010),2.67) (The last decade -- First Half (2011--2015),5.49) (The last decade -- Second Half (2016--2020),3.41)};
\addlegendentry{Automotive}
\addplot[smooth,color=red,mark=diamond*,mark options={scale=1.5},densely dashed] coordinates {(Initial Stages (1990--2002),0.00) (Foundational Years (2003--2005),3.03) (Ramping up (2006--2010),4.00) (The last decade -- First Half (2011--2015),1.10) (The last decade -- Second Half (2016--2020),5.68)};
\addlegendentry{Mobile Systems}
\addplot[smooth,color=teal,mark=*,mark options={scale=1.5},densely dashed] coordinates {(Initial Stages (1990--2002),0.00) (Foundational Years (2003--2005),0.00) (Ramping up (2006--2010),2.67) (The last decade -- First Half (2011--2015),4.40) (The last decade -- Second Half (2016--2020),4.55)};
\addlegendentry{Service-Oriented Systems}
\addplot[smooth,color=orange,mark=square*,mark options={scale=1.5},densely dashed] coordinates {(Initial Stages (1990--2002),0.00) (Foundational Years (2003--2005),0.00) (Ramping up (2006--2010),4.00) (The last decade -- First Half (2011--2015),5.49) (The last decade -- Second Half (2016--2020),2.27)};

\end{axis}
\end{tikzpicture}
\caption{Applications domains over time}
\label{fig:domainsovertime}
\end{figure}
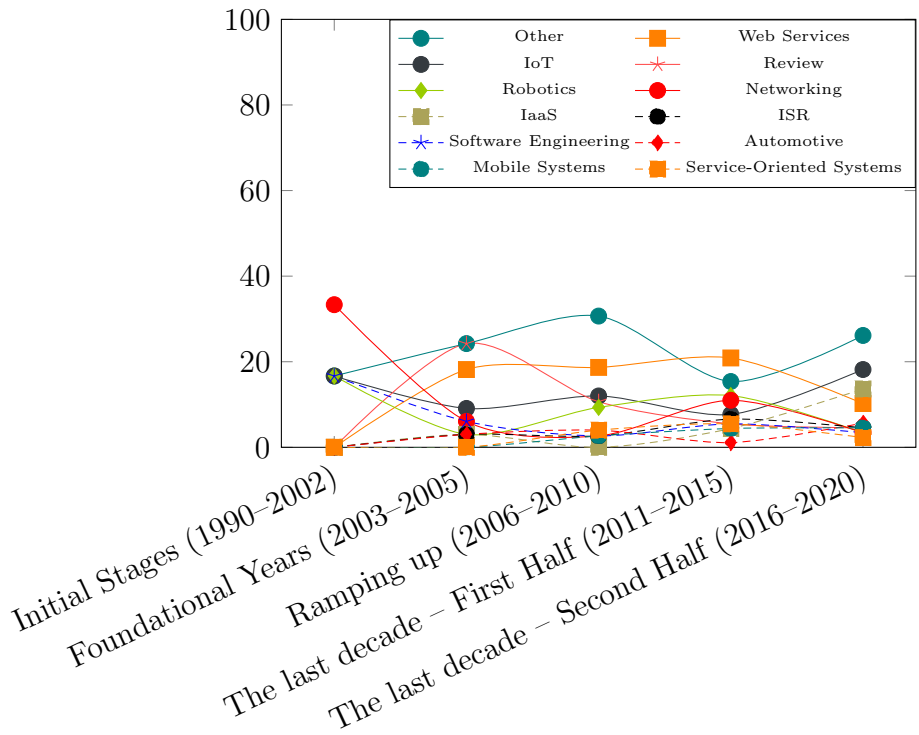

\begin{table}
    \centering
    \scriptsize
    \caption{Summary of attribute matrices. Note that totals can differ from 1.0 when attribute values are unknown or when articles attain multiple values.}
    \begin{tabular}{@{}l@{\hskip 0.4em}l@{\hskip 0.4em}l@{\hskip 0.4em}l@{\hskip 0.4em}l@{\hskip 0.4em}l@{\hskip 0.4em}l@{}}
\toprule    
\textbf{Type} & Algorithm & Architecture & Framework & Mathemat. & Language & Definition \\
\textit{Analytical} & \Chart{0.46} & \Chart{0.03} & \Chart{0.14} & \Chart{0.26} & \Chart{0.06} & \Chart{0.09} \\
\midrule
\textbf{Type} & General arch. & New Framew. & Analysis Techn. & New Pattern & Formal Crit. & New Appr. \\
\textit{Methodology} & \Chart{0.04} & \Chart{0.15} & \Chart{0.08} & \Chart{0.01} & \Chart{0.03} & \Chart{0.70} \\
\midrule
\textbf{Type} & Survey & Review & Evaluation & Reflection & Roadmap & Comparison \\
\textit{Perspective} & \Chart{0.20} & \Chart{0.31} & \Chart{0.04} & \Chart{0.35} & \Chart{0.02} & \Chart{0.08} \\
\midrule
\textbf{Type} & Human & Closed circle \\
\textit{Technological} & \Chart{0.18} & \Chart{0.81} \\
\midrule
\textbf{Formalization} & Yes & No \\
\textit{Analytical} & \Chart{0.94} & \Chart{0.03} \\
\midrule
\textbf{Implementation} & Tool & Model & Framework & Language & Architecture & Algorithm \\
\textit{Technological} & \Chart{0.24} & \Chart{0.23} & \Chart{0.22} & \Chart{0.14} & \Chart{0.10} & \Chart{0.07} \\
\midrule
\textbf{Goals} & Goals & Utility \\
\textit{Technological} & \Chart{0.55} & \Chart{0.25} \\
\midrule
\textbf{Content} & Taxonomy & Challenges & Future Work & Requirements \\
\textit{Perspective} & \Chart{0.14} & \Chart{0.33} & \Chart{0.71} & \Chart{0.16} \\
\midrule
\textbf{Testing} & Design-time & Run-time \\
\textit{Empirical} & \Chart{0.00} & \Chart{0.74} \\
\midrule
\textbf{Strategy} & Monitoring & Adaptation \\
\textit{Empirical} & \Chart{0.83} & \Chart{0.87} \\
\midrule
\textbf{Adaptation} & Parameter & Component \\
\textit{Empirical} & \Chart{0.87} & \Chart{0.09} \\
\midrule
\textbf{Contribution} & Extension & Novel \\
\textit{Analytical} & \Chart{0.31} & \Chart{0.69} \\
\textit{Technological} & \Chart{0.12} & \Chart{0.87} \\
\textit{Methodology} & \Chart{0.09} & \Chart{0.90} \\
\midrule
\textbf{Application} & Case study & Simulated \\ 
\textit{Analytical} & \Chart{0.57} & \Chart{0.40} \\
\textit{Technological} & \Chart{0.60} & \Chart{0.40} \\
\textit{Methodology} & \Chart{0.78} & \Chart{0.20} \\
\textit{Empirical} & \Chart{0.61} & \Chart{0.35} \\
\midrule
\textbf{Evaluation} & Preliminary & Case Study & Industrial & Comparison & Human & Quant. \\
\textit{Analytical} & \Chart{0.09} & \Chart{0.29} & \Chart{0.00} & \Chart{0.17} & \Chart{0.00} & \Chart{0.37} \\
\textit{Technological} & \Chart{0.01} & \Chart{0.35} & \Chart{0.00} & \Chart{0.15} & \Chart{0.00} & \Chart{0.44} \\ 
\textit{Perspective} & \Chart{0.00} & \Chart{0.29} & \Chart{0.00} & \Chart{0.31} & \Chart{0.00} & \Chart{0.16} \\ 
\textit{Methodology} & \Chart{0.00} & \Chart{0.38} & \Chart{0.00} & \Chart{0.16} & \Chart{0.00} & \Chart{0.41} \\ 
\textit{Empirical} & \Chart{0.00} & \Chart{0.39} & \Chart{0.04} & \Chart{0.13} & \Chart{0.00} & \Chart{0.57} \\ 
\bottomrule
    \end{tabular}
    \label{tab:summarytable}
\end{table}

In the following, we provide a chronological overview of research on self-adaptive systems based on our systematic literature review. The complete matrices with the characterizations of all papers are shown at the end of the paper and summarized here:

\begin{itemize}
\item Technological: 81\%~(\PieChart{81}) of papers in this category follow a closed circle approach as opposed to human in the loop, and implementations cover tools (24\%~\PieChart{24}), models (23\%~\PieChart{23}), frameworks (22\%~\PieChart{22}), languages, architectures, and algorithms. 40\%~(\PieChart{40}) of the papers rely on simulations. The vast majority of contributions are novel (87\%~\PieChart{87}), compared to a few extensions of previous work. The methods for evaluation range from quantitative approaches (44\%~\PieChart{44}) and case studies (35\%~\PieChart{35}) to comparisons (15\%~\PieChart{15}).
\item Methodology: The majority of papers in this category (70\%~\PieChart{70}) introduce new approaches for self-adaptive systems, followed by frameworks (15\%~\PieChart{15}), analysis techniques (8\%~\PieChart{8}), and architectures (4\%~\PieChart{4}). The ratio of simulations (20\%~\PieChart{20}) is lower than in the Technological category, and the vast amount of papers provide new contributions (90\%~\PieChart{90}) here as well. Evaluation follows a similar pattern to the Technological category, with quantitative evaluations (41\%~\PieChart{41}) being the most common, followed by case studies (38\%~\PieChart{38}) and comparisons (16\%~\PieChart{16}).
\item Perspective: Many of the papers in this category can be classified as reflections (35\%~\PieChart{35}), reviews (31\%~\PieChart{31}), and surveys (20\%~\PieChart{20}), with papers outlining future work (71\%~\PieChart{71}), challenges (33\%~\PieChart{33}), and requirements (16\%~\PieChart{16}), or providing a taxonomy (14\%~\PieChart{14}). Evaluation methods, if applicable, in this category are mostly focused on comparisons (31\%~\PieChart{31}) and case studies (29\%~\PieChart{29}).
\item Analytical: Many papers in this category focus on algorithms (46\%~\PieChart{46}), followed by mathematical contributions (26\%~\PieChart{26}) and frameworks (14\%~\PieChart{14}). Like other categories, the number of case studies vs.~simulations follows roughly a 60/40 split. A substantial number of papers are extensions of previous work (31\%~\PieChart{31}), and almost all (94\%~\PieChart{94}) papers provide a formalization. In terms of evaluation methods, the largest group are quantitative (37\%~\PieChart{37}), followed by case studies (29\%~\PieChart{29}) and comparisons (17\%~\PieChart{17}).
\item Empirical: Most of the empirical papers focus on run-time (74\%~\PieChart{74}) instead of design-time, with an explicit monitoring (83\%~\PieChart{83}) and/or adaptation (87\%~\PieChart{87}) strategy. Studies again follow the 60/40 split between case studies and simulations. The adaptation techniques focus on parameters (87\%~\PieChart{87}) rather than components (9\%~\PieChart{9}), and evaluation is mostly quantitative (57\%~\PieChart{57}) or done through case studies (39\%~\PieChart{39}).
\end{itemize}

Our analysis revealed the following five time periods, mostly defined by the number of papers published on the topic and their primary research focus:



\begin{itemize}
    \item 1990-2002 - Initial Stages, Section~\ref{sec:1}
    \item 2003-2005 - Foundational Years, Section~\ref{sec:2}
    \item 2006-2010 - Ramping up, Section~\ref{sec:3}
    \item 2011-2015 - The last decade - First Half, Section~\ref{sec:4}
    \item 2016-2020 - The last decade - Second Half, Section~\ref{sec:5}
\end{itemize}

As an orthogonal perspective, the tables at the end of this paper group all 293 papers by category and list their attributes. Table~\ref{tab:summarytable} provides a high-level summary by showing the distribution of attribute values in each category.

We briefly describe the attributes and their values in the following, before characterizing the field in each time period. Note that attributes and their values emerged from our analysis of the papers:

\begin{itemize}
    \item Type: The type of contribution of a paper depends on the paper category. Many of the analytical papers introduce a new algorithm or provide a formal mathematical contribution. Other contribution types of analytical papers include frameworks, definitions, languages, and architectures for self-adaptive systems. The majority of methodology papers introduce a new approach for self-adaptive systems, with other contributions including new frameworks and architectures, new analysis techniques, new patterns, and formal criteria, e.g., for diagnosing components in self-adaptive systems. Perspective papers typically contribute reflections, reviews, or surveys, with a smaller number of papers focusing on evaluating specific frameworks, presenting roadmaps, or comparing approaches to a particular problem. Lastly, the types of contributions of technological papers can be divided into human-in-the-loop and closed-circle approaches, depending on the level of human intervention in the self-adaptive system.
    \item Formalization: The vast majority of analytical papers provides a formalization of their contribution, e.g., using mathematical definitions and formulas.
    \item Implementation: Technological papers provide an implementation of something, ranging from tools, models, and frameworks, to languages, architectures, and algorithms.
    \item Goals: Technological papers can further be distinguished based on whether they focus on the goals of a self-adaptive system and/or associated utility functions. A self-adaptive system should respect the utility while trying to achieve its goal.
    \item Content: The content of the majority of perspective papers is a thorough discussion of future work, with others listing challenges or requirements, or providing a taxonomy.
    \item Testing: While testing can generally be divided into design-time testing and run-time testing depending on whether a system is running during testing, our review only revealed empirical papers focused on run-time testing.
    \item Strategy: Most empirical papers employ an adaptation strategy which defines possible actions and their implementation as well as a monitoring strategy which defines how to extract information from the system.
    \item Adaptation: The majority of empirical papers employ parameter adaptation, i.e., fine tuning of applications through the modification of application variables and deployment parameters, rather than component adaptation, i.e., the replacement, addition, or removal of components.
    \item Contribution: Most papers exclusively present original work, while a minority are extensions of other published work, e.g., journal extensions of conference papers.
    \item Application: The majority of approaches introduced across all paper categories (except perspective) are evaluated using real-world case studies, with less reliance on simulations.
    \item Evaluation: Across all paper categories, the methods for evaluation of research on self-adaptive systems are dominated by quantitative approaches, case studies, and comparisons. We use the attribute value `quantitative' in cases of statistical tests based on simulations, often in absence of a baseline. We use the attribute value `case study' to refer to in-depth studies of one or a small number of systems. We use the attribute value `comparison' to indicate work that was primarily evaluated by comparing to a baseline, e.g., from previous work. Other evaluation methods include real-world (`industrial') and user studies (`human'), or those explicitly labeled by the authors as `preliminary'.
\end{itemize}

In addition, Figures~\ref{fig:papersovertime},~\ref{fig:categoriesovertime}, and~\ref{fig:domainsovertime} visualize the trends of number of papers over time, categories over time, and application domains over time.



\subsection{1990-2002 - Initial Stages}\label{sec:1}

According to our literature review, the earliest reference to self-adaptive systems was in 1990. The period of 1990-2003 was the phase before `The Vision of Autonomic Computing'~\cite{Kephart2003}, was published. This was where the field was in its beginning stages.

References in this phase contained theoretical and model based papers such as a model for dynamic change management~\cite{Kramer1990}, self-stabilizing real-time rule based systems~\cite{Cheng1992}, and convergence for self-stabilizing systems~\cite{Beauguier1999}. These papers were focused on presenting a theoretical model or proving a proposed theorem to advance self-adaptive systems theory. 

By 1998, practical implementations were seen such as architecture based run-time software evolution~\cite{Oreizy1998}, self-supervised category detection~\cite{YAMAUCHI1999}, and self-adaptive control software~\cite{Pham2000}. The papers were among the first to discuss terms like evolution, self-supervision, control theory, and run-time design in the context of self-adaptive systems. Thirty years later, these terms are commonplace and part of the general understanding of the field.

Tables~\ref{tab:1990categories} and~\ref{tab:1990domains} characterize this time period in terms of the prevalent paper categories and application domains.

\begin{table}[t]
\centering
\caption{Categories during initial stages (1990--2002)}
\begin{tabular}{lr}
\toprule
Category & Count \\
\midrule
Technological & 4 \\ 
Analytical & 2 \\ 
\midrule
Total & 6 \\ 
\bottomrule
\end{tabular}
\label{tab:1990categories}
\end{table}

\begin{table}[t]
\centering
\caption{Application domains during initial stages (1990--2002)}
\begin{tabular}{lr}
\toprule
Application Domain & Count \\
\midrule
Networking & 2 \\ 
Software Engineering & 1 \\ 
IoT & 1 \\
Speech Recognition & 1 \\
Robotics & 1 \\
\bottomrule
\end{tabular}
\label{tab:1990domains}
\end{table}

\subsection{2003-2005 - Foundational Years}\label{sec:2}

In the subsequent years after 2002, two seminal works were produced that formed the foundations of self-adaptive systems. In 2003, The Vision of Autonomic Computing~\cite{Kephart2003} was published which kick-started the field of Autonomic Computing. The paper presented a grand challenge of self-configuration, self-optimization, self-healing, and self-protection in computing systems. It foretasted the rising need for these systems in the modern age. Shortly after this, an implementation of these concepts was developed, known as the RAINBOW framework~\cite{Garlan2004}. This framework was architecture based and extensible, meaning existing architectures could be imbued with self-adaptive properties. To this day, the RAINBOW framework is used as a standard, test bed, extensible tool in self-adaptive systems research.

Many papers were published during this time period that were in the perspective category (42\%). As the excitement of a new field grew, many researchers sought to publish their thoughts and ideas as to how the field should and could develop. During this period, there were fewer practical demonstrations and implementations compared to future time periods as the field had to have time to grow and mature. 

This application domain distribution of this period was skewed towards review papers. Of the eight review papers, four were on autonomic computing~\cite{Kephart2003,Kephart2005,Ganek2003,Ganek2004} and four were on self-* properties~\cite{Hales2005,Lemos2005,Babaoglu2005,Cheng2005}. The key messages of these papers were that a new challenge was emerging in the field of autonomic computing and self-* computing. Due to the rise in system complexity, it was necessary to develop a new kind of system that was self-adaptive. These papers also envisioned what such realized systems might look like. 

The most immediate application domain for self-adaptive systems was web services. At the time, technologies such as IaaS and IoT had not fully emerged yet. Web services were a prominent tool used in software engineering. 

The improvement of the system administrator role was a focus area of research by improving collaboration and coordination, rehearsal and planning, maintaining situation awareness, and managing multitasking, interruptions, and diversions~\cite{BARRETT2005}. Utility functions emerged as a potential metric to solve self-adaptive problems~\cite{Walsh2004}. In later years, this would prove to be true. The automatic management of web services was an important test bed to develop self-adaptive algorithms and theory~\cite{Kapoor2005,Houben2005,wolter2004self}. 

There was a major parallel between autonomic computing, self-* computing, and biological systems, and a branch of research was formed to gain inspiration from nature to bring to these systems~\cite{Heylighen2003,Nowostawski2004,Yang2005}. As a precursor to the cloud based systems of the present day, load balancing was an area of research focus. Work forecasting~\cite{Bennani2004}, scheduling algorithms~\cite{Kurmas2004}, and managing system level properties~\cite{White2004} were all part of the groundwork of this research area, as was the placement of replicants in an edge computing scenario~\cite{sivasubramanian2005autonomic}.

IoT and IaaS papers were present during this time. Concepts like autonomous deployment~\cite{Mikic-Rakic2004}, generic control frameworks~\cite{Kandasamy2005}, and self-healing distributed architectures~\cite{Shin2005} were explored, as was the concept of self-aware systems~\cite{andras2004self} in general. These two fields would emerge as strong research fields in later years.

Other research areas include Robotics~\cite{Zhao2005}, Automotive~\cite{DeWolf2005}, Computer~\cite{Sousa2005}, Software Stack~\cite{Wildstrom2005}, Networking~\cite{Gupta2005,Breitgand2005}, Software Engineering~\cite{McKinley2004}, and Intelligence Surveillance Reconnaissance~\cite{Bulusu2004}.

Tables~\ref{tab:2003categories} and~\ref{tab:2003domains} characterize this time period in terms of the prevalent paper categories and application domains, confirming the large number of perspective papers.

\begin{table}[t]
\centering
\caption{Categories during foundational years  (2003--2005)}
\begin{tabular}{lr}
\toprule
Category & Count \\
\midrule
Perspective & 14 \\ 
Technological & 8 \\ 
Methodology & 5 \\ 
Analytical & 3 \\ 
Empirical & 3 \\ 
\midrule
Total & 33 \\ 
\bottomrule
\end{tabular}
\label{tab:2003categories}
\end{table}

\begin{table}[t]
\centering
\caption{Application domains during foundational years (2003--2005)}
\begin{tabular}{lr}
\toprule
Application Domain & Count \\
\midrule
Review & 8 \\ 
Web Services & 6 \\ 
Load Balancing & 3 \\
Bio-inspired & 3\\
IoT & 3 \\
Networking & 2 \\
Software Engineering & 2 \\
\bottomrule
\end{tabular}
\label{tab:2003domains}
\end{table}

\subsection {2006-2010 - Ramping up}\label{sec:3}

The distribution of the years 2006-2010 showed a more even spread across the categories. Technological, Perspective, and Methodology papers were evenly spread. There was still a high demand for perspective type papers but enough time had passed since 2003 for technology and methodology papers to emerge.

Web services was still the highest polling application domain but in this period IoT papers surfaced as the second most frequent domain. 
Web services in this phase focused on the principles and ideas of autonomic and self-adaptive systems developed in the previous phase. These manifested initially in new languages~\cite{webservice0601}, utility functions~\cite{webservice0601}, and distributed solutions as opposed to centralized solutions~\cite{webservice0602}. There was a clear focus on run time instead of design time solutions~\cite{webservice0603, webservice0608} as well as making systems dynamic in their configuration and adaptation~\cite{webservice0604, webservice0606,bhat2006enabling}. Self-healing was an important part of web services research during this time, highlighting the use in workarounds~\cite{webservice0605} and composition cycles~\cite{webservice0607}. The RAINBOW framework was evaluated on a signature case study: Znn.com to assess the effectiveness of self-adaptation with good results~\cite{webservice0609}. In the latter half of this period, the focus of web services grew to become more holistic with concepts such as architectural self-reconfiguration and self-tuning~\cite{webservice0610,webservice0611}, behavioral adaptations~\cite{webservice0612}, expansion into more self-* properties~\cite{webservice0613}, and adaption logic based approaches~\cite{webservice0614}. The emphasis was on adaptation of the entire system as opposed to individual components.

IoT papers were initially focused on solvers~\cite{iot0602, iot0603} with a key driver being dynamic behavior~\cite{iot0601, iot0606}. Fault tolerance~\cite{iot0601} and multi-agent models~\cite{iot0602} were important themes to establish in IoT due to their distributed and always-on nature. A tool called RELAX was developed to handle requirements and uncertainty in a smart home self-adaptive system domain~\cite{iot0604}. Requirements and uncertainty are important in any self-adaptive system, not just IoT systems~\cite{iot0607}. The requirements of the system determine the goals of the adaptations. Appropriately managing uncertainty gives a system flexibility and dynamism. As with web services, a holistic approach to self-adaptation using architectural self-reconfiguration was a part of the research~\cite{iot0605}. Adaption logic in IoT was addressed using heuristics in availability and response time~\cite{iot0608} and in a tool called FUSION~\cite{iot0609}. The research efforts of IoT compared with web services in this phase share a parallel. Initially, solvers and languages were created and tested, then various research concepts were explored like distributed solutions for web services or fault tolerance for IoT. Then, a shift in focus to more holistic approaches to self-adaptability were emphasized such as architectural self-reconfiguration, or the adaptation logic.

This phase was the first time robotics was seen as an application domain with seven papers. Robotics represented a high potential area for the application of self-adaptive systems. By their nature, robots are intended to automatically do the tasks of humans and so self-adaptive robots are a natural progression of the goals. Requirements and modelling~\cite{robotics0601,robotics0602} were the early emphasis with a progression to adaptable software architectures~\cite{robotics0602,robotics0604} and self-organized and distributed systems~\cite{robotics0603}. A reference model to compare adaptation approaches was developed~\cite{robotics0606}. This started to become necessary as the popularity of the field gave rise to multiple different adaptation techniques.  Learning and planning was a component of the research of this phase~\cite{robotics0605,robotics0607}. Using reinforcement learning, planning and architecture self-management was explored~\cite{robotics0605}. The learning research theme was in its early stages in this phase. In future years, learning would prove to be key to widely used tools like machine and deep learning.

This phase also gave rise to nine review papers. Compared to the previous phase, review papers made up a smaller proportion of the total papers in the group. The reviews were conducted on self-organization~\cite{review0601,review0604}, autonomic computing~\cite{review0602, review0605,review0606}, self-healing~\cite{review0603}, and self-adaptation~\cite{review0607, review0608}, as well as a general overview of self-* properties~\cite{berns2009dissecting}. The common theme of this work was to highlight the current work and future needs of the field. One remark confirms the trend highlighted above of needing a more holistic approach to self-adaptation: \textit{"but what is missing is an holistic approach focusing explicitly on providing autonomic properties"}~\cite{review0606}. Written in 2009, it perhaps explains the trend toward holistic solutions to self-adaptive systems in the literature.

This phase showed a variety of other application domains. Decentralized solutions were popular in automated traffic control~\cite{06traffic03}, e-Commerce~\cite{06ecommerce03}, water networks~\cite{06water}, service oriented systems~\cite{06SOS02}, and load balancing solutions~\cite{06load}. Learning was a prominent theme for automotive~\cite{06automotive02}, traffic management control~\cite{06traffic02}, and e-Commerce~\cite{06ecommerce02}. The ability for a system to adapt its adaptation logic was a strong area of focus for mobile systems\cite{06mob01}, ISR~\cite{06ISR02}, e-Commerce~\cite{06ecommerce01}, software engineering applications~\cite{06soft01,06soft02}, and systems on chip~\cite{06SOC}. Evaluation~\cite{06management,06mob02}, self-organization~\cite{06automotive01,06computer01,06computer02}, self-protection~\cite{06programmer}, self-healing~\cite{06automotive03}, resource allocation~\cite{06SOS01}, reflection~\cite{06traffic01}, as well as dynamic solutions~\cite{06ISR01} were all explored during this time. Consistent with the theme of holistic solutions, generic architectures were developed in speech recognition tools~\cite{06speech} as well as generic frameworks such as SASSY~\cite{06SOS03}. Other application domains not listed include UML~\cite{06uml}, Holonic Systems~\cite{06holon}, Video Encoding~\cite{06vid}, task scheduling~\cite{tidwell2008scheduling}, and Physics~\cite{06physics}.

Tables~\ref{tab:2006categories} and~\ref{tab:2006domains} characterize this time period in terms of the prevalent paper categories and application domains, underlining that papers in the `Methodology' category and about Web Services were becoming increasingly dominant.

\begin{table}[t]
\centering
\caption{Categories during ramping up stage (2006--2010)}
\begin{tabular}{lr}
\toprule
Category & Count \\
\midrule
Methodology & 23 \\ 
Technological & 20 \\ 
Perspective & 18 \\ 
Analytical & 8 \\ 
Empirical & 6 \\ 
\midrule
Total & 75 \\ 
\bottomrule
\end{tabular}
\label{tab:2006categories}
\end{table}

\begin{table}[t]
\centering
\caption{Application domains during ramping up stage (2006--2010)}
\begin{tabular}{lr}
\toprule
Application Domain & Count \\
\midrule
Web Services & 14 \\
IoT & 9 \\ 
Review & 8 \\ 
Robotics & 7 \\
E-Commerce & 3\\
Service Oriented Systems & 3 \\
Automated Traffic Control & 3 \\
Automotive & 3 \\
Intelligence Surveillance Reconnaissance & 2 \\
Mobile Systems & 2 \\
Computer & 2 \\
Software Engineering & 2 \\
\bottomrule
\end{tabular}
\label{tab:2006domains}
\end{table}


\subsection {2011-2015 - The last decade - First Half}\label{sec:4}

The first half of the 2010s showed a bimodal result, with technological and methodology papers being the most frequent. The proportion of perspective papers decreased, perhaps attributed to the saturation of them in previous time periods. 

Web services still dominated the domain distribution, followed by robotics and networking. IoT still had a high ranking with seven papers. This was the first time period where IaaS made it in to the top domains. This is in line with the timeline of cloud based services as they rose to prominence.

Web services papers in this time period displayed a slightly different flavor to the previous time periods. Forms of validation were more popular in this phase. Performance~\cite{webservices1101} and integration~\cite{webservices1104} testing, probabilistic model checking~\cite{webservices1105,webservices1116,webservices1113}, quality assurance~\cite{webservices1106}, and evaluation~\cite{webservices1109} indicate that more emphasis was now being put on verifying the outcomes of self-adaptive solutions. It was not enough to claim that a system was self-adaptive, but the claims had to be backed up and tested. Frameworks formed part of the contributions to the research. With the push towards more holistic self-adaptive solutions, this was a natural progression. Multi-model~\cite{webservices1102}, dynamic allocation~\cite{webservices1103}, monitoring~\cite{webservices1107}, and behavioral~\cite{webservies1108} frameworks highlight the variety of holistic generic approaches being explored. The RAINBOW framework still held influence during this period, with a framework called REFRACT extending RAINBOW to target fault avoidance~\cite{webservices1111}. Planning was a key component of this phase -- automatic reconfiguration plans~\cite{webservices1118}, adapting manager optimization~\cite{webservices1115}, and plan generation techniques~\cite{webservices1117} all highlight the continued importance of forecasting in self-adaptive systems. Distributed techniques~\cite{webservices1114}, fault localization~\cite{webservices1112}, and architecture based self-adaptation~\cite{webservices1110} all were continued self-adaptive themes from the previous phase.

Robotics papers also shared an emphasis on building generic frameworks for self-adaptation. Architectural compilers~\cite{robotics1101}, reference models~\cite{robotics1104}, and testing frameworks~\cite{robotics1108} highlight some of the work done here. There was a continued focus on dynamic~\cite{robotics1105} and run-time~\cite{robotics1109} applications as well as verification of systems~\cite{robotics1107}. The modelling of uncertainty became a strong focus during this phase. As decisions in the system are pushed from design time to run time, the amount of possible outcomes for the system dramatically increases. The behavior of the system also becomes non-deterministic. Verification~\cite{robotics1111}, consequence modelling~\cite{robotics1103}, and latency modelling~\cite{webservices1116} all attest to this effort. A strong research theme of the robotics domain in self-adaptive systems is goal modelling. The goals of a robot and associated utility functions are highly important to successful behavior. Dealing with fuzzy goals~\cite{robotics1110}, interactions~\cite{robotics1103}, and learning~\cite{robotics1106} all contributed to this focus.

Networking was a domain that was well represented during this time period. With the explosion of the Internet and always-on devices, networking was a ripe domain to apply self-adaptive principles to. Consistent with the theme of frameworks during this time period, mathematical~\cite{network1101}, scheduling~\cite{network1102}, sensor modelling~\cite{network1103}, and testing~\cite{network1109} frameworks were created. There was a similar emphasis on validation~\cite{network1110} and fault tolerance~\cite{network1108}. The RAINBOW framework was again used as an exemplar during this phase. It was applied to manage and monitor highly populated networks of devices~\cite{network1107}. Consistent with the themes from previous phases, self-organization~\cite{network1104} and self-reconfiguration~\cite{network1105, network1106} were popular research areas in networking.

The IoT domain shares a high overlap with the networking domain as IoT solutions are essentially localized networks. The trends of the IoT research in this phase is consistent with the themes of generic frameworks~\cite{iot1104,iot1105,iot1102,iot1106,mob1103} and validation~\cite{iot1107,iot1103}. 

The IaaS domain entered the top domains in this phase. In the second half of the decade its popularity would explode. During this time, IaaS papers focused on regression testing~\cite{iaas1101}, control theory~\cite{iaas1102}, decision making~\cite{iaas1103,bencomo2013dynamic}, transaction management~\cite{sos1104}, and on benchmarking and elasticity~\cite{iaas1104}. These themes would later on be developed and expanded. 

There were only six reviews in this time period, the lowest proportion of any time period so far. This could be due to the reduced need because of the ongoing work. Reviews in this period focused on self-healing~\cite{review1101,review1104}, self-adaptation~\cite{review1102,review1105}, self-protection~\cite{review1103}, and on control engineering approaches to self-adaptive system design~\cite{control1101}.

Frameworks in ISR~\cite{isr1103,isr1102}, service oriented systems~\cite{sos1102,sos1103,sos1105}, software engineering~\cite{soft1101,soft1103}, and mobile systems~\cite{mob1101,mob1102} highlight the trend of the theme in developing holistic solutions to self-adaptive systems. Continued trends are requirements~\cite{sos1101,iot1101}, dynamic solutions~\cite{soft1104}, multi-agent systems~\cite{isr1105,11resource}, and utility~\cite{isr1101} as well as reliability~\cite{11programmer}, with more abstract ideas like systems evaluation~\cite{villegas2011framework}, uncertainty handling~\cite{robotics1102,isr1106}, and feedback loops~\cite{soft1102,control1102} now getting covered.

Bio-inspired approaches were again present in this phase. These approaches have the understanding that self-adaptive systems are much like biological systems and that there is much inspiration to draw from nature. Papers discuss chemically inspired architectures for reusable models~\cite{bio1101}, as well as cloud based applications inspired by biological principles~\cite{bio1102} and multi-objective control for self-adaptive software design~\cite{mob1104}. These biological inspired approaches are a potential growth area for self-adaptive systems.

Other domain applications not listed include e-Commerce~\cite{11ecommerce}, UML~\cite{11UML}, automotive~\cite{11automotive}, water networks~\cite{11water}, automated traffic management~\cite{11traffic}, fault recovery~\cite{11stack}, video encoding~\cite{11video}, application containers~\cite{soft1105}, and human participation~\cite{isr1104}.

Tables~\ref{tab:2011categories} and~\ref{tab:2011domains} characterize this time period in terms of the prevalent paper categories and application domains, showing a large number of papers in the categories `Technological' and `Methodology', again dominated by the application area of Web Services.

\begin{table}[t]
\centering
\caption{Categories during first half of the last decade (2011--2015)}
\begin{tabular}{lr}
\toprule
Category & Count \\
\midrule
Technological & 37 \\ 
Methodology & 29 \\ 
Empirical & 9 \\ 
Analytical & 8 \\ 
Perspective & 8 \\ 
\midrule
Total & 91 \\ 
\bottomrule
\end{tabular}
\label{tab:2011categories}
\end{table}

\begin{table}[t]
\centering
\caption{Application domains during first half of the last decade (2011--2015)}
\begin{tabular}{lr}
\toprule
Application Domain & Count \\
\midrule
Web Services & 19 \\
Robotics & 11 \\
Networking & 10 \\ 
IoT & 7 \\ 
Intelligence Surveillance Reconnaissance & 6 \\
Service Oriented Systems & 5 \\
Software Engineering & 5 \\
Review & 5 \\
IaaS & 4 \\
Mobile Systems & 4 \\
Bio-inspired & 2 \\
Control Engineering & 2 \\
\bottomrule
\end{tabular}
\label{tab:2011domains}
\end{table}


\subsection {2016-2020 - The last decade - Second Half}\label{sec:5}

The second half of the 2010s displayed a similar distribution to the first half, with technological and methodology papers being most frequent, followed by analytical, perspective, and empirical. This is not surprising as the developments of the field were at a comparable maturation stage.

The distribution of the domains during this period shows an interesting trend. Web services are no longer the most frequent domain, rather IoT and IaaS are the most frequent domains. There is a distinct trend of these cloud based technologies from obscurity (1990-2003) to niche (2003-2010) to growth (2011-2015) to now in the 2020s where cloud based services are mainstream. This trend is reflected in the rise of IoT and IaaS in the domain distribution of the papers.

IoT has become a popular application domain in the last five years of the 2010s. The number of devices with access to the Internet has increased exponentially in quantity but also variety. Devices are not just limited to phones and computers but extend to watches, cars, buildings, sensors, and more. The research in this phase gave rise to a number of exemplars in IoT. Exemplars can be generic such as artifacts or address specific self-adaptive problems. They are used as a demonstration of a working solution in the problem space. This was the period of time where exemplars began to be widely seen and used. DeltaIoT~\cite{iot1605}, an evaluation exemplar and DingNet~\cite{iot1611}, a simulation exemplar highlight the work done. Carrying on from the work done in the previous time periods, frameworks and generic architectures were seen in HAFLoop~\cite{iot1616}, decentralized approaches~\cite{iot1608,iot1607}, and modelling frameworks~\cite{iot1606}. Behavioral modelling of IoT behavior was a continued trend~\cite{iot1613,iot1603,iot1601,sos1601}. Specifically, emergent behaviors were explored~\cite{iot1604}. Emergent behavior is an important concept in self-adaptive systems. These behaviors are the byproduct of allowing decisions to be made at run time. When this occurs, the system may display new behavior not previously conceived or seen before. The appropriate handling of these behaviors is important to a large scale self-adaptive system like IoT. Common research focuses like new languages~\cite{iot1602}, evaluation and testing~\cite{iot1609,raibulet2017evaluation}, modelling~\cite{iot1610}, learning~\cite{iot1612,load1602,camara2016adaptation}, recovery~\cite{beal2016self}, uncertainty~\cite{iot1614}, and integration~\cite{iot1615} were seen in this phase. A general review of self-improving system integration can be found in~\cite{review1602} and industrial experience reports in~\cite{16retro,review1601}.

IaaS became more of a focus during this period. As it went further into the decade, cloud based solutions became more and more common in business and hence in research. It became cheaper to rent out infrastructure in the cloud and outsource maintenance costs than to handle everything in-house. This is also reflected in research focusing on concurrent approaches~\cite{soft1601,vinh2016concurrency} or hierarchical systems~\cite{review1603}.

From the previous time period, the research areas of testing~\cite{iaas1611}, control theory~\cite{iot1608}, decision making~\cite{iaas1603,iaas1604,iaas1609}, and elasticity~\cite{iot1612} in IaaS were expanded on. New areas of research, such as trust~\cite{iaas1601}, structural and parametric adaptation~\cite{iaas1602}, monitoring~\cite{iot1610,sos1602}, modelling~\cite{iaas1606,iaas1607}, and service level maintenance~\cite{iaas1612} were established in the field. Trust in self-adaptive systems is an important concept. Even if the self-adaptation loops are robust and effective, without establishing trust for the system, using these systems in large scale or critical environments is infeasible. This has to to with the understood error rate of the self-adaptive system and the tolerance of the user. In some cases it may be acceptable to have a 20\% error rate in a non-critical scenario but for another critical scenario like a Defence setting, even a 5\% error rate may not be acceptable given the possible consequences. The research efforts in self-adaptive systems are usually split between structural and parametric adaptation. Structural adaptation involves modifying or improving the components of the system whereas parametric adaptation involves optimizing the configurable parameters of the system, leaving the components unchanged. Addressing both of the styles at once is an area of need and potential~\cite{iaas1602}.


The research into web services has benefited from a 20 plus year build up. Consistent with the trend of this time period, exemplars were used to demonstrate the capabilities of self-adaptive web services using TCP communication~\cite{webservices1607}. Multi-agent systems~\cite{webservices1605}, uncertainty~\cite{webservices1603}, planning~\cite{webservices1609}, models~\cite{webservices1608,webservices1606}, and programming concepts~\cite{webservices1601,webservices1602} were all continued research themes into web services. The state of web services after 20 plus years has moved from foundational theory to generic frameworks and to exemplars. Even despite this trend, the various research themes are still being explored and mined for use after 20 years which indicates that there is more to learn.

Cyber-Physical Systems (CPS) are a combination of computation, networking, and physical processes where a physical component is controlled by a chip or software component. This time period is the first time CPS are seen. This indicates that they are a relatively new research area to self-adaptive systems. The dominating application (24\%) to CPS is energy and the dominant adaptation mechanism is MAPE-K~\cite{cyber1601}. A new language, Adaptive CSP was developed to support compositional verification of systems~\cite{cyber1603}. Continuing with the trend of exemplars in this phase, DARTSim represents a simulation of UAVs on a reconnaissance mission communicating via TCP~\cite{cyber1605}. According to~\cite{cyber1606}, a central concept in these systems is homeostasis, the capacity to maintain an operational state despite run-time uncertainty. This is addressed by four principles: collaborative sensing, faulty component isolation from adaptation, enhancing mode switching, and adjusting guards in mode switching. CPS are naturally employed in safety-critical environments as their small nature allows them to be embedded into any physical tool. The successful integration~\cite{cyber1604} and the tracability of these components~\cite{cyber1602} are critical to the field. 

Continuing on with the last time period, bio inspired approaches were seen in this time period, addressing emergent behavior~\cite{bio1601} and artificial DNA~\cite{bio1602}. Security was a focus of this theme with guarantees~\cite{security1601} and verifications~\cite{security1602} being explored. The self-protection aspect of self-adaptive systems has been sprinkled amongst the time periods 
(with works focusing on trust~\cite{yan2006autonomic,wang2006toward,serrano2009trust} and on situational awareness~\cite{yau2006development}), 
however as these systems gain traction and popularity, there will be an increased need to secure these systems in the same fashion as micro-transactions are secured in financial institutions. 
It would not be surprising if in the next decade, security became a more prominent application domain. 

There was a paper in this time period on smart factory or industry 4.0~\cite{16smartfactory}. Seen as the next progression in industrial activity, this application domain has potential to grow going in to the next decade. Exemplars were again seen in this period across other domains using architectural self-healing and self-optimization~\cite{soft1602}.

The frequencies of robotics and networking decreased in this phase. This could be because they are less popular or that there is some overlap between these domains and the top two domains, IoT and IaaS. The second reason is more likely. Planning~\cite{robotics1601,robotics1602,iaas1608,iaas1605}, testing~\cite{robotics1603}, fault tolerance~\cite{network1601}, and uncertainty~\cite{network1603,network1602} all highlight common research trends seen before in the timelines, as is model-predictive control~\cite{load1601}. Mobile systems most likely also share the same similarities with overlap as networking and robotics to IoT and IaaS. They have been a consistent theme across the timelines and have a presence in this one with dynamic decisions~\cite{mob1601}, input space mapping~\cite{mob1602} and emotion measurement~\cite{mob1603}. A review of self-adaptive systems in the context of mobile systems is given in~\cite{mob1604}, one on monitoring self-adaptive applications within edge computing frameworks in~\cite{iaas1610}, and one on learning in self-adaptive systems in~\cite{review1604}.

The automotive application domain increased in this time period compared to the previous time period. This may be explained by the new found viability of smart cars and self-driving cars in recent years. Key trends for this domain were adaptive, scalable, and robust systems. These systems are proactively aware of latency and can act in swarms~\cite{auto1601,auto1602,auto1603,auto1604,auto1605}. 

ISR has had a consistent presence across the timelines. Resilience~\cite{isr1601}, goal theory~\cite{isr1602}, control theory~\cite{isr1603}, and assurance~\cite{isr1604} highlight the research efforts in this time period. ISR is an important application domain to self-adaptive systems. It enables real time situational awareness and allows analysts to make decisions based off current and useful information. In a Defence context, generating this intelligence from data is extremely important to the decision makers.

Other application domains included clonal plasticity~\cite{16clonal}, smart travels~\cite{16smarttravel}, agriculture~\cite{16agri}, UML~\cite{16uml}, system on chip~\cite{16soc}, MAPE-K~\cite{16mape}, traffic management~\cite{16traffic}, holonic systems~\cite{16holon}, and managing support of recoonfigurable software components~\cite{webservices1604}.

Tables~\ref{tab:2016categories} and~\ref{tab:2016domains} characterize this time period in terms of the prevalent paper categories and application domains, underlining the focus on IoT and IaaS in recent years.

\begin{table}[t]
\centering
\caption{Categories during second half of the last decade (2016--2020)}
\begin{tabular}{lr}
\toprule
Category & Count \\
\midrule
Technological & 36 \\ 
Methodology & 22 \\ 
Analytical & 14 \\ 
Perspective & 11 \\ 
Empirical & 5 \\ 
\midrule
Total & 88 \\ 
\bottomrule
\end{tabular}
\label{tab:2016categories}
\end{table}

\begin{table}[t]
\centering
\caption{Application domains during second half of the last decade (2016--2020)}
\begin{tabular}{lr}
\toprule
Application Domain & Count \\
\midrule
IoT & 16 \\
IaaS & 12 \\
Web Services & 9 \\
Cyber Physical Systems & 6 \\
Automotive & 5 \\
Review & 4 \\
Mobile Systems & 4 \\
Intelligence Surveillance Reconnaissance & 4 \\
Robotics & 3 \\
Networking & 3 \\
Load Balancing & 2 \\
Service Oriented Systems & 2 \\
Software Engineering & 2 \\
Security & 2 \\
Bio-inspired & 2 \\
\bottomrule
\end{tabular}
\label{tab:2016domains}
\end{table}

\subsection{Summary}

In this section, we briefly revisit the research questions set out at the beginning of this review to summarize our findings.

RQ1 What is the current state of the art in self-adaptive systems? The current state of the art in self-adaptive systems is focused on developing methodologies and technology in the area of cloud-based services, such as IoT and IaaS. Although research on self-adaptive systems tends to be diverse, empirical and analytical research is currently playing a smaller role, as are other application domains. The importance of self-adaptive systems is rapidly growing in areas such as bio-inspired approaches, security, and cyber physical systems.

RQ2 How has the state of the art evolved over time? In the 1990s, research on self-adaptive systems started with theoretical and model based papers to establish the foundations of the field. Practical implementations and frameworks together with forward-thinking perspective research gave rise to the rapid growth of the field in the 2000s and 2010s, with a need for and a trend towards holistic approaches and exemplars. Throughout the evolution of the field, researchers have published a large number of perspective papers to challenge the status quo and outline the needs of practitioners.

RQ3 Which are the application domains of self-adaptive systems over time? After an initial focus on networking, web services have dominated self-adaptive systems as an application area for much of the field's evolution, up until around 2015 when IoT and IaaS became the most frequent domains. From the beginning, the field has exhibited a large and diverse number of application domains, from robotics and networking to automotive and intelligence surveillance reconnaissance.



\section{Threats to Validity and Limitations}\label{sec:threats}

Unlike related literature reviews on self-adaptive systems which characterize the state of the art of a narrow and specific aspect of self-adaptive systems such as requirements~\cite{yang2014systematic, sucipto2015systematic}, claims and evidence~\cite{review1102}, quality attributes~\cite{mahdavi2017systematic}, and machine learning in the context of SAS~\cite{saputri2020application}, we took a broader view of the literature in this work, which necessarily limits the amount of detail presented for each of the 293 papers. The tables in the appendix provide the high-level overview from our review at a glance.

This systematic literature review was conducted with some assumptions. The dblp database was a suitable database to capture self-adaptive systems. Dblp is a computer science bibliography, and the review would not capture papers outside this bibliography in fields like medicine, science, and other engineering fields. We used the CORE ranking of a publication venue as a proxy for paper quality. The corresponding filtering step may have excluded high-quality papers relevant for our review that were published in other venues.

The application domains mentioned in this paper are subject to the interpretation of the papers. A paper may have multiple application domains but only one was chosen for each paper. This means there is some overlap across the domains. At best, it is useful to get a flavor of the types of papers in self-adaptive systems across the 30 year time period but it is not a comprehensive survey of all the types of domains.

The inclusion and exclusion criteria described in Section~\ref{sec:protocol} bias the selection of primary studies, e.g., by using keywords for pre-filtering. These steps were necessary to handle the large amount of papers. The pre-filtering would likely have resulted in a different set of papers if we had considered abstracts and keywords in addition to titles when computing word frequency. Note that some of these concerns are mitigated by our use of snowballing to pick up papers that were missed through the initial search. Focusing on a single main contribution per paper also introduces bias since papers may have more than one contribution.



\section{Conclusion}

Self-adaptive systems research dates back to the 1990s where theoretical and model based papers established foundational self-adaptive theory. These theories gave rise to practical implementations and frameworks such as the RAINBOW framework in the early 2000s. During this time several perspective papers were published such as the seminal work `The Vision of Autonomic Computing' which outlined the grand challenges of the field moving forward. The ramping up years of 2006-2010 were characterized by principles and ideas leading up to a need for more holistic generic approaches. In the first half of the 2010s the need for holistic generic approaches was met with several new frameworks. By the second half of the 2010s, these frameworks were extended to become exemplars, working solutions with real use cases. In the 2020s, if the popularity of self-adaptive systems continues to grow, these exemplars are likely to turn into mainstream adopted solutions.

The ebb and flow of the application domains across the time period show web services being most popular in the 2000s before IoT and IaaS papers joined them as the most popular in the 2010s. In the late 2010s domains like bio-inspired approaches, security, and cyber physical systems showed promise to grow heading into the 2020s. As time goes on, often an unknown disruptive solution arises that slowly makes its way to the top of the domains. In the 2020s technologies could arise like this from unlikely sources.

In their systematic review on claims and supporting evidence for self-adaptive systems from 2012, Weyns et al.~\cite{review1102} concluded that only a few systematic empirical studies had been undertaken at that point. This trend has not really changed over the last decade: while the overall number of papers on self-adaptive systems continues to grow, less than 8\% of the papers identified in our systematic review focus on the empirical aspect, compared to 36\% technological papers and 27\% methodology papers. Perhaps encouragingly, many approaches are evaluated using real-world case studies, with less reliance on simulations. Going forward, we expect to see a shift towards empirical studies as the research field continues to mature, with industrial case studies in many of the application domains identified here.

For a young research field such as self-adaptive systems, a surprisingly large ratio of papers focus on reflecting on the current state of the field and/or providing a road map going forward (17\% of the papers identified in our review). Due to the size of the field, very few of these perspective papers encompass self-adaptive systems as a whole, instead focusing on particular sub-classes of or challenges related to self-adaptive systems. In contrast, we provide a high-level overview of the field across categories and application domains.

In our future work, we aim to work towards closing some of the gaps identified in this systematic literature review, with a particular focus on systematic empirical studies. In line with Gerostathopoulos et al.'s recent study~\cite{gerostathopoulos2021we} which concluded that ``most data of users and the environment used in experiments is synthetically generated'', we aim to experiment with human subjects -- an aspect that has not received much attention from the self-adaptive systems research community thus far. At the same time, the maturity of the field now allows for the development and deployment of such systems in real environments, with large-scale evaluations using the empirical methods that are well-established in other areas of software engineering~\cite{ralph2020empirical}. Through our industry collaborators, we further will put particular focus on self-adaptation for cyber-physical systems. Cyber-physical systems have to handle uncertainty and change during operation, control their emergent behavior, and be scalable and tolerant to threats~\cite{cyber1601}, yet their complexities introduce new challenges to self-adaptive systems that are difficult to capture without real-world evaluations.





\begin{table*}
\centering
\caption{Analytical\label{tab:analytical}
}\def\arraystretch{0}\setlength{\tabcolsep}{2.0pt}%
\small\hspace*{-28mm}%
\begin{tabular}{ll|llllll|ll|ll|ll|lllllll|l} 
\toprule
 & & \multicolumn{6}{|c|}{Type} & \multicolumn{2}{|c|}{Applic.} & \multicolumn{2}{|c|}{Contr.} &  \multicolumn{2}{|c|}{Formal.} & \multicolumn{7}{|c|}{Evaluation Method} \\ 
\midrule
\rot{Reference} & \rot{Year} & \rot{Algorithm} & \rot{Architecture} & \rot{Framework} & \rot{Mathematical} & \rot{Language} & \rot{Definition} & \rot{Case study} & \rot{Simulated} & \rot{Extension} & \rot{Novel} & \rot{Yes} & \rot{No} & \rot{Preliminary} & \rot{Case Study} & \rot{Industrial} & \rot{Comparison} & \rot{Human Subject} & \rot{Quantitative} & \rot{Unknown/None} & \rot{Domain} \\
\midrule
\cite{Cheng1992} & 1992 & \OK &  &  &  &  &  &  & \OK &  & \OK & \OK &  &  &  &  &  &  & \OK &  & Networking \\
\cite{Beauguier1999} & 1999 &  &  &  & \OK &  &  &  & \OK &  & \OK &  &  &  &  &  &  &  & \OK &  & Networking \\
\cite{Breitgand2005} & 2005 & \OK &  &  &  &  &  & \OK &  &  & \OK & \OK &  &  & \OK &  &  &  &  &  & Networking \\
\cite{wolter2004self} & 2005 & \OK &  &  &  &  &  & \OK &  &  & \OK & \OK &  &  & \OK &  &  &  &  &  & Web Services \\
\cite{Zhao2005} & 2005 &  &  & \OK &  &  &  & \OK &  &  & \OK & \OK &  &  & \OK &  &  &  &  &  & Robotics \\
\cite{wang2006toward} & 2006 &  &  & \OK &  &  &  &  & \OK &  & \OK & \OK &  &  &  &  & \OK &  &  &  & Security \\
\cite{06holon} & 2007 &  &  & \OK &  &  &  & \OK &  &  & \OK & \OK &  &  & \OK &  &  &  &  &  & Holonic Systems \\
\cite{webservice0603} & 2008 & \OK &  &  &  &  &  & \OK &  & \OK &  & \OK &  &  &  &  & \OK &  &  &  & Web Services \\
\cite{06physics} & 2009 &  &  &  & \OK &  &  & \OK &  & \OK &  & \OK &  & \OK &  &  &  &  &  &  & Physics \\
\cite{berns2009dissecting} & 2009 &  &  &  &  &  & \OK &  &  &  & \OK & \OK &  &  &  &  &  &  &  & \OK & Other \\
\cite{iot0603} & 2009 &  &  &  & \OK &  &  & \OK &  & \OK &  & \OK &  & \OK &  &  &  &  &  &  & IoT \\
\cite{06ecommerce02} & 2010 &  &  &  &  &  & \OK & \OK &  & \OK &  & \OK &  &  & \OK &  &  &  &  &  & e-Commerce \\
\cite{webservice0612} & 2010 & \OK &  &  &  &  &  & \OK &  &  & \OK & \OK &  &  & \OK &  &  &  &  &  & Web Services \\
\cite{bio1101} & 2011 &  &  &  & \OK &  &  & \OK &  & \OK &  & \OK &  &  &  &  &  &  & \OK &  & Bio-inspired \\
\cite{iot1101} & 2011 &  &  &  &  &  & \OK & \OK &  &  & \OK & \OK &  &  &  &  &  &  & \OK &  & IoT \\
\cite{network1101} & 2011 &  &  &  & \OK &  &  & \OK &  & \OK &  & \OK &  &  &  &  &  &  &  & \OK & Networking \\
\cite{soft1101} & 2011 &  &  &  & \OK &  &  & \OK &  &  & \OK & \OK &  &  & \OK &  &  &  &  &  & Software Engineering \\
\cite{sos1101} & 2011 &  &  &  & \OK &  &  &  & \OK &  & \OK & \OK &  &  &  &  &  &  & \OK &  & Service-Oriented Systems \\
\cite{isr1102} & 2012 &  &  &  &  & \OK &  & \OK &  &  & \OK & \OK &  &  & \OK &  &  &  &  &  & ISR \\
\cite{network1103} & 2012 &  & \OK &  &  &  &  & \OK &  &  & \OK & \OK &  &  &  &  & \OK &  &  &  & Networking \\
\cite{robotics1106} & 2013 & \OK &  &  &  &  &  &  & \OK &  & \OK & \OK &  &  &  &  &  &  & \OK &  & Robotics \\
\cite{beal2016self} & 2016 & \OK &  &  &  &  &  &  & \OK & \OK &  & \OK &  &  &  &  &  &  & \OK &  & IoT \\
\cite{bio1601} & 2016 & \OK &  &  &  &  &  &  & \OK & \OK &  & \OK &  & \OK &  &  &  &  &  &  & Bio-inspired \\
\cite{iaas1601} & 2016 & \OK &  &  &  &  &  & \OK &  &  & \OK & \OK &  &  &  &  &  &  & \OK &  & IaaS \\
\cite{network1601} & 2016 & \OK &  &  &  &  &  &  & \OK &  & \OK & \OK &  &  &  &  &  &  & \OK &  & Networking \\
\cite{vinh2016concurrency} & 2016 &  &  &  & \OK &  &  &  & \OK &  & \OK & \OK &  &  &  &  &  &  &  & \OK & Other \\
\cite{iaas1605} & 2017 &  &  &  & \OK &  &  &  & \OK & \OK &  & \OK &  &  &  &  & \OK &  &  &  & IaaS \\
\cite{iot1604} & 2017 & \OK &  &  &  & \OK &  & \OK &  &  & \OK & \OK &  &  & \OK &  &  &  &  &  & IoT \\
\cite{auto1603} & 2018 & \OK &  &  &  &  &  & \OK &  &  & \OK & \OK &  &  &  &  &  &  & \OK &  & Automotive \\
\cite{iaas1609} & 2018 & \OK &  &  &  &  &  &  & \OK & \OK &  &  & \OK &  &  &  & \OK &  &  &  & IaaS \\
\cite{iot1610} & 2018 & \OK &  &  &  &  &  &  & \OK &  & \OK & \OK &  &  &  &  &  &  & \OK &  & IoT \\
\cite{robotics1601} & 2018 & \OK &  &  &  &  &  & \OK &  &  & \OK & \OK &  &  & \OK &  &  &  &  &  & Robotics \\
\cite{robotics1602} & 2018 &  &  & \OK &  &  &  & \OK &  &  & \OK & \OK &  &  &  &  &  &  & \OK &  & Robotics \\
\cite{security1601} & 2018 &  &  & \OK &  &  &  &  & \OK & \OK &  & \OK &  &  &  &  & \OK &  &  &  & Security \\
\cite{webservices1609} & 2019 & \OK &  &  &  &  &  &  & \OK &  & \OK & \OK &  &  &  &  &  &  & \OK &  & Web Services \\
\bottomrule
\end{tabular}
\end{table*}

\begin{table*}
\centering
\label{tab:technological1}
\centering\vspace{-59mm}%
\caption{Technological (1/3)}\def\arraystretch{0}
\setlength{\tabcolsep}{2.pt}
\small\hspace*{-30mm}%
\begin{tabular}{ll|ll|llllll|ll|ll|ll|lllllll|l} 
\toprule
& & \multicolumn{2}{|c|}{Type} & \multicolumn{6}{|c|}{Implementation} & \multicolumn{2}{|c|}{Applic.} & \multicolumn{2}{|c|}{Contr.} & \multicolumn{2}{|c|}{Goals} & \multicolumn{7}{|c|}{Evaluation Method} \\ 
\midrule
\rot{Reference} & \rot{Year} & \rot{Human in the loop} & \rot{Closed circle} & \rot{Tool} & \rot{Model} & \rot{Framework} & \rot{Language} & \rot{Architecture} & \rot{Algorithm} & \rot{Case Study} & \rot{Simulated} & \rot{Extension} & \rot{Novel} & \rot{Goals} & \rot{Utility} & \rot{Preliminary} & \rot{Case Study} & \rot{Industrial} & \rot{Comparison} & \rot{Human Subject} & \rot{Quantitative} & \rot{Unknown/None} & \rot{Domain} \\
\midrule
\cite{Kramer1990} & 1990 &  & \OK &  & \OK &  &  &  &  & \OK &  &  & \OK &  &  &  & \OK &  &  &  &  &  & Software Engineering \\
\cite{Oreizy1998} & 1998 &  & \OK & \OK &  &  &  &  &  &  & \OK &  & \OK &  &  &  &  &  &  &  & \OK &  & IoT \\
\cite{YAMAUCHI1999} & 1999 &  & \OK &  &  &  &  &  & \OK &  & \OK &  & \OK &  &  &  &  &  &  &  & \OK &  & Speech Recognition \\
\cite{Pham2000} & 2000 &  & \OK &  &  & \OK &  &  &  & \OK &  &  & \OK & \OK &  &  & \OK &  &  &  &  &  & Robotics \\
\cite{Garlan2004} & 2004 & \OK &  &  &  &  &  & \OK &  & \OK &  &  & \OK & \OK & \OK &  & \OK &  &  &  &  &  & Software Engineering \\
\cite{Kurmas2004} & 2004 &  & \OK & \OK &  &  &  &  &  &  & \OK & \OK &  &  &  &  &  &  & \OK &  &  &  & Load Balancing \\
\cite{Mikic-Rakic2004} & 2004 &  & \OK &  &  &  &  & \OK &  &  & \OK & \OK &  & \OK &  &  &  &  &  &  & \OK &  & IaaS \\
\cite{Nowostawski2004} & 2004 &  & \OK &  &  & \OK &  &  &  &  & \OK &  & \OK & \OK &  &  &  &  &  &  & \OK &  & Bio-inspired \\
\cite{Walsh2004} & 2004 &  & \OK &  &  &  &  & \OK &  & \OK &  &  & \OK & \OK & \OK &  &  &  &  &  & \OK &  & Web Services \\
\cite{Houben2005} & 2005 & \OK &  & \OK &  &  &  &  &  & \OK &  &  & \OK &  &  &  & \OK &  &  &  &  &  & Web Services \\
\cite{Wildstrom2005} & 2005 &  & \OK &  &  & \OK &  &  &  &  & \OK &  & \OK &  &  &  &  &  &  &  & \OK &  & Software Stack \\
\cite{Yang2005} & 2005 &  & \OK & \OK &  &  &  &  &  &  & \OK &  & \OK &  &  &  &  &  & \OK &  &  &  & Bio-inspired \\
\cite{06ecommerce01} & 2006 &  & \OK &  &  &  & \OK &  &  & \OK &  &  & \OK &  & \OK &  & \OK &  &  &  &  &  & e-Commerce \\
\cite{bhat2006enabling} & 2006 &  & \OK &  & \OK &  &  &  &  &  & \OK & \OK &  & \OK & \OK &  &  &  &  &  & \OK &  & Networking \\
\cite{webservice0601} & 2006 & \OK &  &  &  &  & \OK &  &  &  & \OK &  & \OK &  & \OK &  &  &  &  &  & \OK &  & Web Services \\
\cite{yan2006autonomic} & 2006 &  & \OK &  &  & \OK &  &  &  &  & \OK &  & \OK &  &  &  &  &  &  &  & \OK &  & Security \\
\cite{06ISR02} & 2007 &  & \OK & \OK &  &  &  &  &  &  & \OK &  & \OK & \OK &  &  &  &  &  &  &  & \OK & ISR \\
\cite{06vid} & 2007 &  & \OK & \OK &  &  &  &  &  & \OK &  &  & \OK &  &  &  & \OK &  &  &  &  &  & Video Encoding \\
\cite{webservice0602} & 2007 &  & \OK &  &  & \OK &  &  &  &  & \OK & \OK &  &  &  &  &  &  &  &  &  & \OK & Web Services \\
\cite{06load} & 2008 & \OK &  &  & \OK &  &  &  &  &  & \OK &  & \OK & \OK &  &  &  &  &  &  &  & \OK & Load Balancing \\
\cite{iot0601} & 2008 &  & \OK &  & \OK &  &  &  &  &  & \OK &  & \OK & \OK &  &  &  &  &  &  & \OK &  & IoT \\
\cite{robotics0601} & 2008 &  & \OK &  &  & \OK &  &  &  & \OK &  &  & \OK & \OK &  &  &  &  &  &  &  & \OK & Robotics \\
\cite{06uml} & 2009 &  & \OK &  & \OK &  &  &  &  & \OK &  &  & \OK & \OK &  &  & \OK &  &  &  &  &  & UML \\
\cite{iot0604} & 2009 &  & \OK &  &  &  & \OK &  &  & \OK &  &  & \OK & \OK &  &  & \OK &  &  &  &  &  & IoT \\
\cite{iot0605} & 2009 & \OK &  &  &  &  &  & \OK &  & \OK &  &  & \OK & \OK &  &  & \OK &  &  &  &  &  & IoT \\
\cite{robotics0604} & 2009 &  & \OK &  &  &  &  & \OK &  & \OK &  &  & \OK & \OK &  &  & \OK &  &  &  &  &  & Robotics \\
\cite{webservice0607} & 2009 & \OK &  &  &  & \OK &  &  &  & \OK &  &  & \OK &  &  &  &  &  & \OK &  &  &  & Web Services \\
\cite{webservice0608} & 2009 &  & \OK &  & \OK &  &  &  &  & \OK &  &  & \OK & \OK &  &  &  &  &  &  & \OK &  & Web Services \\
\cite{06ecommerce03} & 2010 &  & \OK &  &  &  &  & \OK &  & \OK &  &  & \OK &  &  &  & \OK &  &  &  &  &  & e-Commerce \\
\cite{iot0606} & 2010 &  & \OK &  &  & \OK &  &  &  & \OK &  &  & \OK & \OK & \OK &  & \OK &  &  &  &  &  & IoT \\
\cite{robotics0606} & 2010 &  & \OK &  &  &  &  &  &  & \OK &  &  & \OK & \OK &  &  & \OK &  &  &  &  &  & Robotics \\
\cite{robotics0607} & 2010 & \OK &  &  &  &  & \OK &  &  & \OK &  &  & \OK & \OK &  &  & \OK &  &  &  &  &  & Robotics \\
\cite{11programmer} & 2011 &  & \OK &  & \OK &  &  &  &  &  & \OK &  & \OK &  &  &  &  &  &  &  & \OK &  & Programmer \\
\cite{11resource} & 2011 &  & \OK &  & \OK &  &  &  &  &  & \OK &  & \OK & \OK &  &  &  &  &  &  & \OK &  & Resource Management \\
\cite{11stack} & 2011 &  & \OK &  &  &  & \OK &  &  &  & \OK &  & \OK &  &  &  &  &  &  &  &  & \OK & Software Stack \\
\cite{iot1102} & 2011 &  & \OK &  & \OK &  &  &  &  &  & \OK &  & \OK & \OK & \OK &  &  &  & \OK &  &  &  & IoT \\
\cite{iot1103} & 2011 &  & \OK &  &  &  &  &  & \OK &  & \OK &  & \OK &  & \OK &  &  &  &  &  & \OK &  & IoT \\
\cite{robotics1101} & 2011 &  & \OK & \OK &  &  &  &  &  &  & \OK &  & \OK & \OK &  &  &  &  &  &  & \OK &  & Robotics \\
\cite{sos1102} & 2011 &  & \OK &  &  & \OK &  &  &  &  & \OK & \OK &  &  &  &  &  &  &  &  & \OK &  & Service-Oriented Systems \\
\bottomrule
\end{tabular}
\end{table*}

\begin{table*}
\centering
\label{tab:technological2}
\centering\vspace{-59mm}%
\caption{Technological (2/3)}\def\arraystretch{0}
\setlength{\tabcolsep}{2.pt}
\small\hspace*{-30mm}%
\begin{tabular}{ll|ll|llllll|ll|ll|ll|lllllll|l} 
\toprule
& & \multicolumn{2}{|c|}{Type} & \multicolumn{6}{|c|}{Implementation} & \multicolumn{2}{|c|}{Applic.} & \multicolumn{2}{|c|}{Contr.} & \multicolumn{2}{|c|}{Goals} & \multicolumn{7}{|c|}{Evaluation Method} \\ 
\midrule
\rot{Reference} & \rot{Year} & \rot{Human in the loop} & \rot{Closed circle} & \rot{Tool} & \rot{Model} & \rot{Framework} & \rot{Language} & \rot{Architecture} & \rot{Algorithm} & \rot{Case Study} & \rot{Simulated} & \rot{Extension} & \rot{Novel} & \rot{Goals} & \rot{Utility} & \rot{Preliminary} & \rot{Case Study} & \rot{Industrial} & \rot{Comparison} & \rot{Human Subject} & \rot{Quantitative} & \rot{Unknown/None} & \rot{Domain} \\
\midrule
\cite{sos1103} & 2011 &  & \OK &  &  &  &  & \OK &  &  & \OK &  & \OK & \OK &  &  &  &  &  &  & \OK &  & Service-Oriented Systems \\
\cite{webservices1101} & 2011 &  & \OK &  &  & \OK &  &  &  &  & \OK &  & \OK & \OK &  &  &  &  & \OK &  &  &  & Web Services \\
\cite{webservices1102} & 2011 &  & \OK &  &  & \OK &  &  &  & \OK &  &  & \OK & \OK &  &  & \OK &  &  &  &  &  & Web Services \\
\cite{webservices1103} & 2011 &  & \OK &  &  & \OK &  &  &  & \OK &  &  & \OK &  &  &  & \OK &  &  &  &  &  & Web Services \\
\cite{bio1102} & 2012 &  & \OK &  &  & \OK &  &  &  &  & \OK &  & \OK &  &  &  &  &  &  &  & \OK &  & Bio-inspired \\
\cite{iot1104} & 2012 &  & \OK &  &  & \OK &  &  &  & \OK &  &  & \OK & \OK &  &  & \OK &  &  &  &  &  & IoT \\
\cite{isr1103} & 2012 &  & \OK &  & \OK &  &  &  &  & \OK &  &  & \OK & \OK &  &  & \OK &  &  &  &  &  & ISR \\
\cite{network1104} & 2012 &  & \OK &  &  &  &  & \OK &  & \OK &  &  & \OK & \OK &  &  &  &  &  &  & \OK &  & Networking \\
\cite{robotics1104} & 2012 &  & \OK &  & \OK &  &  &  &  & \OK &  &  & \OK & \OK &  &  & \OK &  &  &  &  &  & Robotics \\
\cite{soft1102} & 2012 &  & \OK &  &  &  & \OK &  &  &  & \OK &  & \OK &  &  & \OK &  &  &  &  &  &  & Software Engineering \\
\cite{soft1103} & 2012 & \OK &  &  &  &  & \OK &  &  & \OK &  &  & \OK &  & \OK &  &  &  &  &  & \OK &  & Software Engineering \\
\cite{11automotive} & 2013 &  & \OK &  &  &  & \OK &  &  & \OK &  &  & \OK &  &  &  & \OK &  &  &  &  &  & Automotive \\
\cite{bencomo2013dynamic} & 2013 &  & \OK &  & \OK &  &  &  &  & \OK &  & \OK &  & \OK & \OK &  &  &  &  &  & \OK &  & Other \\
\cite{mob1101} & 2013 &  & \OK &  &  & \OK &  &  &  & \OK &  &  & \OK & \OK &  &  & \OK &  &  &  &  &  & Mobile Systems \\
\cite{robotics1107} & 2013 &  & \OK &  &  &  & \OK &  &  & \OK &  & \OK &  &  &  &  &  &  &  &  & \OK &  & Robotics \\
\cite{soft1105} & 2013 & \OK &  & \OK &  &  &  &  &  &  & \OK &  & \OK &  &  &  &  &  & \OK &  &  &  & Software Engineering \\
\cite{webservices1107} & 2013 &  & \OK &  &  & \OK &  &  &  & \OK &  &  & \OK & \OK &  &  &  &  & \OK &  &  &  & Web Services \\
\cite{iot1105} & 2014 &  & \OK &  &  &  & \OK &  &  & \OK &  &  & \OK &  &  &  & \OK &  &  &  &  &  & IoT \\
\cite{mob1103} & 2014 &  & \OK &  &  &  &  &  & \OK & \OK &  &  & \OK &  &  &  &  &  &  &  & \OK &  & Mobile Systems \\
\cite{robotics1108} & 2014 &  &  &  &  &  &  &  & \OK &  & \OK &  & \OK & \OK &  &  &  &  &  &  & \OK &  & Robotics \\
\cite{robotics1109} & 2014 & \OK &  &  & \OK &  &  &  &  & \OK &  &  & \OK & \OK & \OK &  & \OK &  &  &  &  &  & Robotics \\
\cite{robotics1110} & 2014 &  & \OK &  & \OK &  &  &  &  & \OK &  &  & \OK & \OK &  &  & \OK &  &  &  &  &  & Robotics \\
\cite{webservices1110} & 2014 &  & \OK &  & \OK &  &  &  &  & \OK &  & \OK &  & \OK &  &  &  &  &  &  & \OK &  & Web Services \\
\cite{webservices1111} & 2014 &  & \OK &  &  & \OK &  &  &  & \OK &  & \OK &  &  & \OK &  &  &  &  &  & \OK &  & Web Services \\
\cite{isr1104} & 2015 & \OK &  &  &  & \OK &  &  &  & \OK &  &  & \OK & \OK & \OK &  &  &  & \OK &  &  &  & ISR \\
\cite{isr1105} & 2015 &  & \OK &  &  &  &  &  & \OK & \OK &  &  & \OK & \OK &  &  &  &  & \OK &  &  &  & ISR \\
\cite{network1108} & 2015 & \OK &  & \OK &  &  &  &  &  &  & \OK &  & \OK & \OK &  &  &  &  &  &  & \OK &  & Networking \\
\cite{webservices1114} & 2015 &  & \OK &  &  &  & \OK &  &  & \OK &  &  & \OK &  &  &  & \OK &  &  &  &  &  & Web Services \\
\cite{webservices1115} & 2015 &  & \OK & \OK &  &  &  &  &  & \OK &  &  & \OK & \OK &  &  & \OK &  &  &  &  &  & Web Services \\
\cite{webservices1116} & 2015 &  & \OK &  & \OK &  &  &  &  &  & \OK &  & \OK & \OK & \OK &  &  &  &  &  & \OK &  & Web Services \\
\cite{16smartfactory} & 2016 &  & \OK &  &  & \OK &  &  &  & \OK &  &  & \OK &  &  &  & \OK &  &  &  &  &  & Smart Factory \\
\cite{iot1602} & 2016 &  & \OK &  &  &  & \OK &  &  & \OK &  & \OK &  & \OK &  &  & \OK &  &  &  &  &  & IoT \\
\cite{iot1603} & 2016 &  & \OK & \OK &  &  &  &  &  & \OK &  &  &  & \OK &  &  & \OK &  &  &  &  &  & IoT \\
\cite{isr1601} & 2016 &  & \OK &  &  & \OK &  &  &  & \OK &  &  & \OK &  &  &  & \OK &  &  &  &  &  & ISR \\
\cite{mob1601} & 2016 &  & \OK &  &  & \OK &  &  &  &  & \OK &  & \OK &  & \OK &  &  &  &  &  & \OK &  & Mobile Systems \\
\cite{network1602} & 2016 &  & \OK &  &  &  &  &  & \OK & \OK &  &  & \OK &  &  &  & \OK &  &  &  &  &  & Networking \\
\cite{soft1601} & 2016 &  & \OK &  &  & \OK &  &  &  &  & \OK &  & \OK &  &  &  &  &  &  &  & \OK &  & Software Engineering \\
\cite{webservices1601} & 2016 &  & \OK & \OK &  &  &  &  &  & \OK &  &  & \OK & \OK &  &  & \OK &  &  &  &  &  & Web Services \\
\cite{webservices1602} & 2016 &  & \OK &  & \OK &  &  &  &  &  & \OK &  & \OK & \OK & \OK &  &  &  & \OK &  &  &  & Web Services \\
\bottomrule
\end{tabular}
\end{table*}

\begin{table*}
\centering
\label{tab:technological3}
\centering\vspace{-30mm}%
\caption{Technological (3/3)}\def\arraystretch{0}\setlength{\tabcolsep}{2.7pt}
\small\hspace*{-34mm}%
\begin{tabular}{ll|ll|llllll|ll|ll|ll|lllllll|l} 
\toprule
& & \multicolumn{2}{|c|}{Type} & \multicolumn{6}{|c|}{Implementation} & \multicolumn{2}{|c|}{Applic.} & \multicolumn{2}{|c|}{Contr.} & \multicolumn{2}{|c|}{Goals} & \multicolumn{7}{|c|}{Evaluation Method} \\ 
\midrule
\rot{Reference} & \rot{Year} & \rot{Human in the loop} & \rot{Closed circle} & \rot{Tool} & \rot{Model} & \rot{Framework} & \rot{Language} & \rot{Architecture} & \rot{Algorithm} & \rot{Case Study} & \rot{Simulated} & \rot{Extension} & \rot{Novel} & \rot{Goals} & \rot{Utility} & \rot{Preliminary} & \rot{Case Study} & \rot{Industrial} & \rot{Comparison} & \rot{Human Subject} & \rot{Quantitative} & \rot{Unknown/None} & \rot{Domain} \\
\midrule
\cite{16mape} & 2017 &  & \OK &  &  &  &  & \OK &  &  & \OK &  & \OK & \OK & \OK &  &  &  & \OK &  &  &  & MAPE-K \\
\cite{16smarttravel} & 2017 &  & \OK &  &  &  & \OK &  &  &  & \OK &  & \OK & \OK &  &  &  &  & \OK &  &  &  & Smart Traveller \\
\cite{auto1601} & 2017 &  & \OK & \OK &  &  &  &  &  & \OK &  &  & \OK &  &  &  &  &  &  &  & \OK &  & Automotive \\
\cite{iaas1606} & 2017 &  & \OK &  & \OK &  &  &  &  & \OK &  &  & \OK & \OK &  &  &  &  &  &  & \OK &  & IaaS \\
\cite{iot1605} & 2017 & \OK &  & \OK &  &  &  &  &  & \OK &  &  & \OK & \OK &  &  &  &  & \OK &  &  &  & IoT \\
\cite{iot1606} & 2017 &  & \OK & \OK &  &  &  &  &  & \OK &  &  & \OK &  &  &  &  &  & \OK &  &  &  & IoT \\
\cite{iot1607} & 2017 &  & \OK & \OK &  &  &  &  &  &  & \OK &  & \OK &  &  &  &  &  &  &  & \OK &  & IoT \\
\cite{iot1608} & 2017 &  & \OK &  &  & \OK &  &  &  & \OK &  &  & \OK &  & \OK &  &  &  &  &  & \OK &  & IoT \\
\cite{isr1603} & 2017 &  & \OK & \OK &  &  &  &  &  & \OK &  &  & \OK & \OK &  &  & \OK &  &  &  &  &  & ISR \\
\cite{load1601} & 2017 &  & \OK &  & \OK &  &  &  &  &  & \OK &  & \OK & \OK & \OK &  &  &  & \OK &  &  &  & Load Balancing \\
\cite{sos1602} & 2017 & \OK &  & \OK &  &  &  &  &  & \OK &  & \OK &  &  &  &  &  &  &  &  & \OK &  & Service-Oriented Systems \\
\cite{webservices1604} & 2017 & \OK &  &  &  &  & \OK &  &  & \OK &  & \OK &  &  &  &  &  &  &  &  & \OK &  & Web Services \\
\cite{webservices1605} & 2017 &  & \OK & \OK &  &  &  &  &  & \OK &  &  & \OK & \OK &  &  &  &  &  &  & \OK &  & Web Services \\
\cite{webservices1606} & 2017 &  & \OK &  & \OK &  &  &  &  & \OK &  &  & \OK &  &  &  & \OK &  &  &  &  &  & Web Services \\
\cite{16traffic} & 2018 &  & \OK &  & \OK &  &  &  &  & \OK &  &  & \OK & \OK & \OK &  & \OK &  &  &  &  &  & Traffic Mgmt. \\
\cite{auto1604} & 2018 &  & \OK &  & \OK &  &  &  &  & \OK &  &  & \OK & \OK &  &  & \OK &  &  &  &  &  & Automotive \\
\cite{cyber1603} & 2018 &  & \OK & \OK &  &  &  &  &  & \OK &  &  & \OK &  &  &  &  &  &  &  & \OK &  & Cyber Physical Systems \\
\cite{soft1602} & 2018 & \OK &  & \OK &  &  &  &  &  &  & \OK &  & \OK & \OK & \OK &  &  &  &  &  & \OK &  & Software Engineering \\
\cite{webservices1607} & 2018 &  & \OK & \OK &  &  &  &  &  &  & \OK &  & \OK &  & \OK &  &  &  &  &  & \OK &  & Web Services \\
\cite{16agri} & 2019 &  & \OK & \OK &  &  &  &  &  &  & \OK &  & \OK & \OK & \OK &  &  &  &  &  & \OK &  & Agriculture \\
\cite{cyber1605} & 2019 & \OK &  & \OK &  &  &  &  &  &  & \OK &  & \OK &  &  &  &  &  & \OK &  &  &  & Cyber Physical Systems \\
\cite{iaas1612} & 2019 &  & \OK &  &  &  &  &  & \OK &  & \OK &  & \OK &  &  &  &  &  &  &  & \OK &  & IaaS \\
\cite{iot1611} & 2019 & \OK &  & \OK &  &  &  &  &  & \OK &  &  & \OK & \OK &  &  &  &  &  &  & \OK &  & IoT \\
\cite{iot1612} & 2019 &  & \OK &  & \OK &  &  &  &  & \OK &  &  & \OK & \OK & \OK &  &  &  &  &  & \OK &  & IoT \\
\cite{iot1613} & 2019 &  & \OK &  &  &  & \OK &  &  & \OK &  & \OK &  &  & \OK &  & \OK &  &  &  &  &  & IoT \\
\cite{load1602} & 2019 &  & \OK &  & \OK &  &  &  &  &  & \OK &  & \OK &  &  &  &  &  &  &  & \OK &  & Load Balancing \\
\cite{iot1616} & 2020 & \OK &  &  &  &  &  & \OK &  & \OK &  &  & \OK & \OK & \OK &  &  &  &  &  & \OK &  & IoT \\
\bottomrule
\end{tabular}
\end{table*}

\begin{table*}
\centering
\label{tab:perspective}\vspace{-33mm}
\caption{Perspective}\def\arraystretch{0}\setlength{\tabcolsep}{3.pt}%
\small\hspace*{-22mm}%
\begin{tabular}{ll|llllll|llll|lllllll|l} 
\toprule
 & & \multicolumn{6}{|c|}{Type} & \multicolumn{4}{|c|}{Content} & \multicolumn{7}{|c|}{Evaluation Method} \\ 
\midrule
\rot{Reference} & \rot{Year} & \rot{Survey} & \rot{Review} & \rot{Evaluation of framework} & \rot{Reflection} & \rot{Roadmap} & \rot{Comparison} & \rot{Taxonomy} & \rot{Challenges} & \rot{Future Work} & \rot{Requirements} & \rot{Preliminary} & \rot{Case Study} & \rot{Industrial} & \rot{Comparison} & \rot{Human Subject} & \rot{Quantitative} & \rot{Unknown/None} & \rot{Domain} \\
\midrule
\cite{Ganek2003} & 2003 &  &  &  & \OK &  &  &  & \OK & \OK & \OK &  &  &  &  &  &  & \OK & Review \\
\cite{Heylighen2003} & 2003 &  &  &  & \OK &  &  &  &  & \OK &  &  &  &  &  &  &  & \OK & Bio-inspired \\
\cite{Kephart2003} & 2003 &  &  &  & \OK &  &  &  & \OK & \OK & \OK &  &  &  &  &  &  & \OK & Review \\
\cite{andras2004self} & 2004 &  &  &  & \OK &  &  &  &  & \OK &  &  &  &  &  &  &  & \OK & IoT \\
\cite{BARRETT2005} & 2004 &  &  &  & \OK &  &  &  &  & \OK &  &  & \OK &  &  &  &  &  & Web Services \\
\cite{Ganek2004} & 2004 &  &  &  & \OK &  &  &  & \OK &  &  &  &  &  &  &  &  & \OK & Review \\
\cite{Gupta2005} & 2004 &  &  &  & \OK &  &  & \OK &  &  &  &  & \OK &  &  &  &  &  & Networking \\
\cite{Hales2005} & 2004 &  &  &  & \OK &  &  &  &  & \OK &  &  & \OK &  &  &  &  &  & Review \\
\cite{Lemos2005} & 2004 &  &  &  & \OK &  &  & \OK &  &  &  &  &  &  &  &  &  & \OK & Review \\
\cite{McKinley2004} & 2004 &  &  &  & \OK &  &  &  & \OK & \OK & \OK &  &  &  &  &  &  & \OK & Software Engineering \\
\cite{Babaoglu2005} & 2005 &  & \OK &  &  &  &  &  &  & \OK &  &  &  &  &  &  &  & \OK & Review \\
\cite{Cheng2005} & 2005 &  & \OK &  &  &  &  &  &  & \OK & \OK &  & \OK &  &  &  &  &  & Review \\
\cite{Kephart2005} & 2005 & \OK &  &  &  &  &  &  & \OK &  &  &  &  &  &  &  &  & \OK & Review \\
\cite{Sousa2005} & 2005 &  &  &  & \OK &  &  &  &  &  & \OK &  &  &  &  &  & \OK &  & Computer \\
\cite{review0601} & 2006 &  &  &  & \OK &  &  &  &  & \OK &  &  &  &  &  &  &  & \OK & Review \\
\cite{review0602} & 2006 & \OK &  &  &  &  &  &  &  & \OK &  &  &  &  &  &  &  & \OK & Review \\
\cite{review0603} & 2007 & \OK &  &  &  &  &  &  &  & \OK &  &  & \OK &  &  &  &  &  & Review \\
\cite{06soft01} & 2008 &  &  &  &  &  & \OK &  &  & \OK &  &  &  &  & \OK &  &  &  & Software Engineering \\
\cite{review0604} & 2008 &  & \OK &  &  &  &  &  &  & \OK &  &  &  &  &  &  & \OK &  & Review \\
\cite{review0605} & 2008 & \OK &  &  &  &  &  &  & \OK &  & \OK &  &  &  &  &  &  & \OK & Review \\
\cite{06soft02} & 2009 &  & \OK &  &  &  &  &  &  & \OK &  &  &  &  &  &  & \OK &  & Software Engineering \\
\cite{06traffic01} & 2009 &  &  &  & \OK &  &  &  & \OK &  &  &  & \OK &  &  &  &  &  & Traffic Mgmt. \\
\cite{review0606} & 2009 & \OK &  &  &  &  &  &  &  & \OK &  &  &  &  & \OK &  &  &  & Review \\
\cite{review0607} & 2009 &  & \OK &  &  &  &  &  & \OK & \OK &  &  &  &  & \OK &  &  &  & Review \\
\cite{serrano2009trust} & 2009 &  & \OK &  &  &  &  &  & \OK & \OK &  &  &  &  &  &  & \OK &  & Security \\
\cite{webservice0609} & 2009 &  &  & \OK &  &  &  &  &  & \OK &  &  & \OK &  &  &  &  &  & Web Services \\
\cite{06management} & 2010 &  & \OK &  &  &  &  &  &  & \OK &  &  &  &  & \OK &  &  &  & Management \\
\cite{06traffic03} & 2010 &  &  &  & \OK &  &  &  & \OK &  &  &  & \OK &  &  &  &  &  & Traffic Mgmt. \\
\cite{iot0607} & 2010 &  &  &  & \OK &  &  &  &  &  & \OK &  &  &  &  &  & \OK &  & IoT \\
\cite{iot0608} & 2010 &  &  &  & \OK &  &  &  &  & \OK &  &  & \OK &  &  &  &  &  & IoT \\
\cite{review0608} & 2010 &  &  &  &  & \OK &  &  & \OK &  &  &  &  &  & \OK &  &  &  & Review \\
\cite{webservice0613} & 2010 &  &  &  & \OK &  &  &  &  &  & \OK &  & \OK &  &  &  &  &  & Web Services \\
\cite{review1101} & 2011 & \OK &  &  &  &  &  &  &  & \OK &  &  &  &  & \OK &  &  &  & Review \\
\cite{control1101} & 2012 & \OK &  &  &  &  &  & \OK &  &  &  &  &  &  &  &  & \OK &  & Control Engineering \\
\cite{review1102} & 2012 &  & \OK &  &  &  &  &  &  & \OK &  &  &  &  & \OK &  &  &  & Review \\
\cite{review1103} & 2012 & \OK &  &  &  &  &  & \OK &  &  &  &  &  &  & \OK &  &  &  & Review \\
\cite{network1107} & 2013 &  &  & \OK &  &  &  &  &  & \OK &  &  & \OK &  &  &  &  &  & Networking \\
\cite{review1104} & 2015 & \OK &  &  &  &  &  &  &  & \OK &  &  &  &  & \OK &  &  &  & Review \\
\cite{review1105} & 2015 & \OK &  &  &  &  &  & \OK & \OK & \OK &  &  &  &  & \OK &  &  &  & Review \\
\cite{webservices1117} & 2015 &  &  &  &  &  & \OK &  & \OK & \OK &  &  &  &  & \OK &  &  &  & Web Services \\
\cite{16holon} & 2016 &  & \OK &  &  &  &  &  &  & \OK &  &  & \OK &  &  &  &  &  & Holonic Systems \\
\cite{cyber1601} & 2016 &  & \OK &  &  &  &  &  &  & \OK &  &  &  &  & \OK &  &  &  & Cyber Physical Systems \\
\cite{iaas1602} & 2016 &  &  &  &  &  & \OK & \OK &  & \OK &  &  &  &  & \OK &  &  &  & IaaS \\
\cite{raibulet2017evaluation} & 2017 &  &  &  & \OK &  &  &  &  & \OK &  &  &  &  & \OK &  &  &  & Software Engineering \\
\cite{16uml} & 2018 &  & \OK &  &  &  &  &  &  & \OK &  &  &  &  & \OK &  &  &  & UML \\
\cite{iaas1610} & 2018 &  & \OK &  &  &  &  &  & \OK & \OK &  &  &  &  & \OK &  &  &  & IaaS \\
\cite{review1601} & 2018 &  &  &  &  &  & \OK & \OK &  & \OK &  &  & \OK &  &  &  &  &  & Review \\
\cite{review1602} & 2018 &  & \OK &  &  &  &  &  & \OK &  &  &  &  &  &  &  & \OK &  & Review \\
\cite{review1603} & 2018 &  & \OK &  &  &  &  &  &  & \OK &  &  & \OK &  &  &  &  &  & Review \\
\cite{mob1604} & 2019 &  & \OK &  &  &  &  &  & \OK & \OK &  &  &  &  &  &  & \OK &  & Mobile Systems \\
\cite{review1604} & 2019 &  & \OK &  &  &  &  &  & \OK &  &  &  & \OK &  &  &  &  &  & Review \\
\bottomrule
\end{tabular}
\end{table*}

\begin{table*}
\centering\vspace{-43mm}%
\label{tab:methodology1}
\caption{Methodology (1/2)}\def\arraystretch{0}\setlength{\tabcolsep}{3.3pt}
\small\hspace*{-28mm}%
\begin{tabular}{ll|llllll|ll|ll|lllllll|l} 
\toprule
 & & \multicolumn{6}{|c|}{Type} & \multicolumn{2}{|c|}{Applic.} & \multicolumn{2}{|c|}{Contr.} & \multicolumn{7}{|c|}{Evaluation Method} \\ 
\midrule
\rot{Reference} & \rot{Year} & \rot{General architecture} & \rot{New Framework} & \rot{Analysis Technique} & \rot{New Pattern} & \rot{Formal Criteria} & \rot{New Approach} & \rot{Case study} & \rot{Simulated} & \rot{Extension} & \rot{Novel} & \rot{Preliminary} & \rot{Case Study} & \rot{Industrial} & \rot{Comparison} & \rot{Human Subject} & \rot{Quantitative} & \rot{Unknown/None} & \rot{Domain} \\
\midrule
\cite{White2004} & 2004 & \OK &  &  &  &  &  &  & \OK &  & \OK &  &  &  &  &  & \OK &  & Load Balancing \\
\cite{DeWolf2005} & 2005 &  &  &  &  &  & \OK & \OK &  &  & \OK &  &  &  &  &  &  &  & Automotive \\
\cite{Kandasamy2005} & 2005 &  & \OK &  &  &  &  & \OK &  &  & \OK &  &  &  &  &  & \OK &  & IoT \\
\cite{Kapoor2005} & 2005 &  &  &  &  &  & \OK & \OK &  &  & \OK &  & \OK &  &  &  &  &  & Web Services \\
\cite{Shin2005} & 2005 &  &  &  &  &  & \OK & \OK &  &  & \OK &  &  &  &  &  &  &  & IoT \\
\cite{06ISR01} & 2006 &  &  &  &  &  & \OK & \OK &  &  & \OK &  & \OK &  &  &  &  &  & ISR \\
\cite{06mob01} & 2006 &  &  &  &  &  & \OK & \OK &  &  & \OK &  & \OK &  &  &  & \OK &  & Mobile Systems \\
\cite{06speech} & 2006 & \OK &  &  &  &  &  & \OK &  &  & \OK &  &  &  &  &  &  & \OK & Speech Recognition \\
\cite{yau2006development} & 2006 &  & \OK &  &  &  &  & \OK &  &  & \OK &  &  &  & \OK &  &  &  & Security \\
\cite{06programmer} & 2007 &  &  & \OK &  &  &  & \OK &  &  & \OK &  &  &  &  &  & \OK &  & Programmer \\
\cite{06computer01} & 2008 &  &  &  &  &  & \OK & \OK &  &  & \OK &  & \OK &  &  &  &  &  & Computer \\
\cite{06computer02} & 2008 &  & \OK &  &  &  &  & \OK &  &  & \OK &  &  &  &  &  & \OK &  & Computer \\
\cite{iot0602} & 2008 &  &  &  &  &  & \OK &  & \OK & \OK &  &  &  &  & \OK &  &  &  & IoT \\
\cite{robotics0602} & 2008 &  &  &  &  &  & \OK &  & \OK &  & \OK &  &  &  &  &  &  & \OK & Robotics \\
\cite{robotics0603} & 2008 &  &  &  &  &  & \OK &  & \OK &  & \OK &  &  &  & \OK &  &  &  & Robotics \\
\cite{tidwell2008scheduling} & 2008 &  &  & \OK &  &  &  &  & \OK &  & \OK &  &  &  &  &  & \OK &  & Networking \\
\cite{webservice0604} & 2008 &  &  &  &  &  & \OK & \OK &  & \OK &  &  &  &  &  &  & \OK &  & Web Services \\
\cite{webservice0605} & 2008 &  &  &  &  &  & \OK & \OK &  &  & \OK &  & \OK &  &  &  &  &  & Web Services \\
\cite{webservice0606} & 2008 &  &  &  &  &  & \OK &  & \OK &  & \OK &  &  &  &  &  & \OK &  & Web Services \\
\cite{06SOC} & 2009 &  &  &  &  &  & \OK &  & \OK &  & \OK &  &  &  &  &  & \OK &  & System on Chip \\
\cite{06SOS02} & 2009 &  &  &  &  &  & \OK & \OK &  &  & \OK &  &  &  & \OK &  &  &  & Service-Oriented Systems \\
\cite{06SOS03} & 2009 &  & \OK &  &  &  &  & \OK &  &  & \OK &  & \OK &  &  &  &  &  & Service-Oriented Systems \\
\cite{robotics0605} & 2009 &  &  &  &  &  & \OK & \OK &  &  & \OK &  & \OK &  &  &  &  &  & Robotics \\
\cite{webservice0610} & 2009 &  &  &  &  &  & \OK & \OK &  &  & \OK &  & \OK &  &  &  &  &  & Web Services \\
\cite{06automotive03} & 2010 &  &  &  &  &  & \OK & \OK &  &  & \OK &  & \OK &  &  &  &  &  & Automotive \\
\cite{06mob02} & 2010 &  &  &  &  &  & \OK & \OK &  &  & \OK &  &  &  & \OK &  &  &  & Mobile Systems \\
\cite{iot0609} & 2010 &  & \OK &  &  &  &  & \OK &  &  & \OK &  &  &  &  &  & \OK &  & IoT \\
\cite{webservice0614} & 2010 &  &  &  &  &  & \OK &  & \OK &  & \OK &  &  &  &  &  & \OK &  & Web Services \\
\cite{iaas1101} & 2011 &  &  &  &  &  & \OK &  & \OK &  & \OK &  &  &  &  &  & \OK &  & IaaS \\
\cite{network1102} & 2011 &  & \OK &  &  &  &  & \OK &  &  & \OK &  &  &  &  &  & \OK &  & Networking \\
\cite{robotics1102} & 2011 &  &  &  &  &  & \OK & \OK &  &  & \OK &  &  &  &  &  & \OK &  & Robotics \\
\cite{robotics1103} & 2011 &  &  &  &  &  & \OK & \OK &  &  & \OK &  &  &  &  &  & \OK &  & Robotics \\
\cite{villegas2011framework} & 2011 &  & \OK &  &  &  &  & \OK &  & \OK &  &  &  &  &  &  & \OK &  & Other \\
\cite{webservices1104} & 2011 &  &  &  &  &  & \OK & \OK &  &  & \OK &  & \OK &  &  &  &  &  & Web Services \\
\cite{network1105} & 2012 &  &  &  &  &  & \OK & \OK &  &  & \OK &  &  &  &  &  & \OK &  & Networking \\
\cite{robotics1105} & 2012 &  & \OK &  &  &  &  &  & \OK &  & \OK &  &  &  &  &  & \OK &  & Robotics \\
\cite{soft1104} & 2012 &  &  &  &  &  & \OK & \OK &  &  & \OK &  &  &  & \OK &  &  &  & Software Engineering \\
\cite{sos1104} & 2012 &  &  &  & \OK &  &  & \OK &  &  & \OK &  &  &  &  &  & \OK &  & Service-Oriented Systems \\
\cite{webservices1105} & 2012 &  &  &  &  &  & \OK & \OK &  &  &  &  &  &  & \OK &  &  &  & Web Services \\
\cite{webservices1106} & 2012 &  &  &  &  &  & \OK &  & \OK &  & \OK &  &  &  &  &  & \OK &  & Web Services \\
\cite{11UML} & 2013 &  &  &  &  &  & \OK & \OK &  & \OK &  &  & \OK &  &  &  &  &  & UML \\
\cite{iaas1102} & 2013 &  &  &  &  &  & \OK & \OK &  &  & \OK &  & \OK &  &  &  &  &  & IaaS \\
\cite{mob1102} & 2013 &  & \OK &  &  &  &  & \OK &  &  & \OK &  & \OK &  &  &  &  &  & Mobile Systems \\
\cite{sos1105} & 2013 &  &  & \OK &  &  &  &  & \OK &  & \OK &  &  &  &  &  & \OK &  & Service-Oriented Systems \\
\cite{webservices1109} & 2013 &  &  &  &  &  & \OK & \OK &  &  & \OK &  &  &  & \OK &  &  &  & Web Services \\
\bottomrule
\end{tabular}
\end{table*}

\begin{table*}
\centering\vspace{-43mm}%
\label{tab:methodology2}
\caption{Methodology (2/2)}\def\arraystretch{0}\setlength{\tabcolsep}{3.3pt}
\small\hspace*{-28mm}%
\begin{tabular}{ll|llllll|ll|ll|lllllll|l} 
\toprule
 & & \multicolumn{6}{|c|}{Type} & \multicolumn{2}{|c|}{Applic.} & \multicolumn{2}{|c|}{Contr.} & \multicolumn{7}{|c|}{Evaluation Method} \\ 
\midrule
\rot{Reference} & \rot{Year} & \rot{General architecture} & \rot{New Framework} & \rot{Analysis Technique} & \rot{New Pattern} & \rot{Formal Criteria} & \rot{New Approach} & \rot{Case study} & \rot{Simulated} & \rot{Extension} & \rot{Novel} & \rot{Preliminary} & \rot{Case Study} & \rot{Industrial} & \rot{Comparison} & \rot{Human Subject} & \rot{Quantitative} & \rot{Unknown/None} & \rot{Domain} \\
\midrule
\cite{webservies1108} & 2013 &  & \OK &  &  &  &  & \OK &  &  & \OK &  & \OK &  &  &  &  &  & Web Services \\
\cite{11video} & 2014 &  &  &  &  &  & \OK & \OK &  &  & \OK &  & \OK &  &  &  &  &  & Video Encoding \\
\cite{iot1106} & 2014 &  &  &  &  &  & \OK & \OK &  &  & \OK &  & \OK &  &  &  &  &  & IoT \\
\cite{robotics1111} & 2014 &  &  &  &  &  & \OK & \OK &  &  & \OK &  &  &  &  &  & \OK &  & Robotics \\
\cite{webservices1112} & 2014 &  &  &  &  & \OK &  &  &  & \OK &  &  &  &  & \OK &  &  &  & Web Services \\
\cite{webservices1113} & 2014 &  &  & \OK &  &  &  & \OK &  &  & \OK &  & \OK &  &  &  &  &  & Web Services \\
\cite{iot1107} & 2015 &  & \OK &  &  &  &  & \OK &  &  & \OK &  &  &  &  &  & \OK &  & IoT \\
\cite{isr1106} & 2015 &  &  &  &  &  & \OK & \OK &  &  & \OK &  & \OK &  &  &  &  &  & ISR \\
\cite{mob1104} & 2015 &  &  &  &  &  & \OK & \OK &  &  & \OK &  &  &  &  &  & \OK &  & Mobile Systems \\
\cite{network1109} & 2015 &  & \OK &  &  &  &  & \OK &  &  & \OK &  &  &  & \OK &  &  &  & Networking \\
\cite{webservices1116} & 2015 &  &  &  &  &  & \OK & \OK &  &  & \OK &  &  &  &  &  &  &  & Web Services \\
\cite{webservices1118} & 2015 &  &  &  &  &  & \OK &  & \OK &  & \OK &  &  &  &  &  & \OK &  & Web Services \\
\cite{bio1602} & 2016 &  &  &  &  &  & \OK & \OK &  &  & \OK &  &  &  &  &  & \OK &  & Bio-inspired \\
\cite{camara2016adaptation} & 2016 &  &  &  &  &  & \OK & \OK &  & \OK &  &  & \OK &  &  &  &  &  & Web Services \\
\cite{cyber1602} & 2016 &  &  &  &  &  & \OK & \OK &  &  & \OK &  &  &  &  &  & \OK &  & Cyber Physical Systems \\
\cite{iaas1603} & 2016 &  &  &  &  &  & \OK & \OK &  &  & \OK &  &  &  &  &  & \OK &  & IaaS \\
\cite{iaas1604} & 2016 &  &  &  &  &  & \OK &  & \OK &  & \OK &  &  &  &  &  & \OK &  & IaaS \\
\cite{isr1602} & 2016 &  &  &  &  &  & \OK & \OK &  &  & \OK &  & \OK &  &  &  &  &  & ISR \\
\cite{mob1602} & 2016 &  &  &  &  &  & \OK & \OK &  &  & \OK &  &  &  & \OK &  &  &  & Mobile Systems \\
\cite{sos1601} & 2016 &  &  &  &  &  & \OK &  & \OK &  & \OK &  &  &  & \OK &  &  &  & Service-Oriented Systems \\
\cite{16soc} & 2017 &  &  &  &  &  & \OK & \OK &  &  & \OK &  &  &  &  &  & \OK &  & System on Chip \\
\cite{auto1602} & 2017 &  &  &  &  &  & \OK &  & \OK &  & \OK &  &  &  & \OK &  &  &  & Automotive \\
\cite{iaas1607} & 2017 &  &  &  &  &  & \OK & \OK &  &  & \OK &  & \OK &  &  &  &  &  & IaaS \\
\cite{iot1609} & 2017 &  &  &  &  &  & \OK & \OK &  &  & \OK &  & \OK &  &  &  &  &  & IoT \\
\cite{16clonal} & 2018 &  &  & \OK &  &  &  & \OK &  &  & \OK &  &  &  &  &  & \OK &  & Clonal Plasticity \\
\cite{16retro} & 2018 &  &  &  &  &  & \OK & \OK &  &  & \OK &  &  &  &  &  & \OK &  & Retrofitting systems \\
\cite{cyber1604} & 2018 &  &  &  &  &  & \OK & \OK &  &  & \OK &  & \OK &  &  &  &  &  & Cyber Physical Systems \\
\cite{isr1604} & 2018 &  &  &  &  & \OK &  & \OK &  &  & \OK &  & \OK &  &  &  &  &  & ISR \\
\cite{network1603} & 2018 &  &  &  &  &  & \OK & \OK &  &  & \OK &  & \OK &  &  &  &  &  & Networking \\
\cite{security1602} & 2018 &  &  &  &  &  & \OK & \OK &  &  & \OK &  & \OK &  &  &  &  &  & Security \\
\cite{webservices1608} & 2018 &  &  &  &  &  & \OK & \OK &  &  & \OK &  & \OK &  &  &  &  &  & Web Services \\
\cite{cyber1606} & 2019 &  &  & \OK &  &  &  & \OK &  &  & \OK &  & \OK &  &  &  &  &  & Cyber Physical Systems \\
\cite{iot1614} & 2019 &  &  &  &  &  & \OK & \OK &  &  & \OK &  & \OK &  &  &  &  &  & IoT \\
\cite{iot1615} & 2019 & \OK &  &  &  &  &  & \OK &  & \OK &  &  & \OK &  &  &  &  &  & IoT \\
\bottomrule
\end{tabular}
\end{table*}

\begin{table*}
\centering
\caption{Empirical\label{tab:empirical}
}\def\arraystretch{0}\setlength{\tabcolsep}{1pt}%
\small\hspace*{-30mm}%
\begin{tabular}{ll|ll|ll|ll|ll|lllllll|l} 
\toprule
 & & \multicolumn{2}{|c|}{Testing} & \multicolumn{2}{|c|}{Strategy} & \multicolumn{2}{|c|}{Applic.} & \multicolumn{2}{|c|}{\begin{tabular}[x]{@{}c@{}}Adaptation\\Technique\end{tabular}} & \multicolumn{7}{|c|}{\begin{tabular}[x]{@{}c@{}}Evaluation\\Method\end{tabular}} \\ 
\midrule
\rot{Reference} & \rot{Year} & \rot{Design-time} & \rot{Run-time} & \rot{Monitoring} & \rot{Adaptation} & \rot{Case Study} & \rot{Simulated} & \rot{Parameter} & \rot{Component} & \rot{Preliminary} & \rot{Case Study} & \rot{Industrial} & \rot{Comparison} & \rot{Human Subject} & \rot{Quantitative} & \rot{Unknown/None} & \rot{Domain} \\
\midrule
\cite{Bennani2004} & 2004 &  &  & \OK &  &  & \OK & \OK &  &  &  &  &  &  & \OK &  & Load Balancing \\
\cite{Bulusu2004} & 2004 &  & \OK &  & \OK & \OK &  & \OK &  &  &  &  &  &  & \OK &  & ISR \\
\cite{sivasubramanian2005autonomic} & 2005 &  & \OK & \OK & \OK &  & \OK &  &  &  &  &  &  &  & \OK &  & Web Services \\
\cite{06SOS01} & 2006 &  & \OK & \OK & \OK & \OK &  & \OK & \OK &  & \OK &  &  &  &  &  & Service-Oriented Systems \\
\cite{06automotive01} & 2009 &  & \OK & \OK & \OK & \OK &  & \OK &  &  & \OK &  &  &  &  &  & Automotive \\
\cite{06automotive02} & 2009 &  & \OK &  & \OK &  & \OK & \OK &  &  &  &  &  &  & \OK &  & Automotive \\
\cite{06traffic02} & 2009 &  & \OK & \OK & \OK & \OK &  & \OK &  &  &  &  & \OK &  &  &  & Traffic Mgmt. \\
\cite{webservice0611} & 2009 &  &  & \OK &  & \OK &  & \OK &  &  & \OK &  &  &  &  &  & Web Services \\
\cite{06water} & 2010 &  & \OK & \OK & \OK &  &  & \OK &  &  &  &  &  &  & \OK &  & Water Networks \\
\cite{isr1101} & 2011 &  &  & \OK & \OK & \OK &  &  & \OK &  & \OK &  &  &  &  &  & ISR \\
\cite{11ecommerce} & 2012 &  & \OK & \OK & \OK & \OK &  & \OK &  &  & \OK &  &  &  & \OK &  & e-Commerce \\
\cite{control1102} & 2012 &  & \OK & \OK & \OK & \OK &  & \OK &  &  & \OK &  &  &  &  &  & Control Engineering \\
\cite{network1106} & 2012 &  &  &  & \OK &  & \OK & \OK &  &  &  &  &  &  & \OK &  & Networking \\
\cite{11traffic} & 2013 &  & \OK & \OK & \OK &  & \OK & \OK &  &  &  &  &  &  & \OK &  & Traffic Mgmt. \\
\cite{11water} & 2013 &  &  & \OK & \OK &  & \OK & \OK &  &  &  &  &  &  & \OK &  & Water Networks \\
\cite{iaas1103} & 2014 &  & \OK & \OK & \OK & \OK &  & \OK &  &  & \OK &  & \OK &  & \OK &  & IaaS \\
\cite{iaas1104} & 2015 &  & \OK & \OK & \OK & \OK &  & \OK &  &  & \OK &  &  &  &  &  & IaaS \\
\cite{network1110} & 2015 &  & \OK & \OK & \OK & \OK &  & \OK &  &  &  &  &  &  & \OK &  & Networking \\
\cite{iaas1608} & 2017 &  & \OK & \OK & \OK & \OK &  & \OK &  &  &  & \OK &  &  &  &  & IaaS \\
\cite{mob1603} & 2017 &  & \OK & \OK & \OK & \OK &  & \OK &  &  &  &  &  &  & \OK &  & Mobile Systems \\
\cite{auto1605} & 2018 &  &  &  &  &  & \OK &  &  &  &  &  &  &  & \OK &  & Automotive \\
\cite{iaas1611} & 2018 &  & \OK & \OK & \OK &  & \OK & \OK &  &  &  &  & \OK &  &  &  & IaaS \\
\cite{robotics1603} & 2018 &  & \OK & \OK & \OK & \OK &  & \OK &  &  & \OK &  &  &  &  &  & Robotics \\
\bottomrule
\end{tabular}
\end{table*}








\end{document}